\documentclass[usenatbib,usegraphicx]{mn2e}
\bibpunct{(}{)}{;}{a}{}{,}
\usepackage{txfonts}

\newcommand{\dd}{$\rm deg^{2}$}
\newcommand{\flux}{$\rm erg \, s^{-1} \, cm^{-2}$}

\begin{document}

\title[The XMM-LSS Catalogue I]
{The XMM-LSS catalogue: X-ray sources and associated optical data.
Version I}
\author[M. Pierre et al]{M. Pierre$^{1}$\thanks{E-mail:
mpierre@cea.fr },  L. Chiappetti$^{2}$, F. Pacaud$^{1}$,  A.
Gueguen$^{1}$, C. Libbrecht$^{1}$,
\newauthor B.~Altieri$^3$, H.
Aussel$^1$, P.~Gandhi$^4$\thanks{Present address: RIKEN Cosmic
Radiation Lab, 2-1 Hirosawa, Wakoshi, Saitama, 351-0198, Japan},
O.~Garcet$^5$, E.~Gosset$^5$, L.~Paioro$^2$, T.J.~Ponman$^6$,
\newauthor A.M.~Read$^7$, A.~Refregier$^{1}$, J.-L.~Starck$^8$$^{,1}$,
J.~Surdej$^5$, I.~Valtchanov$^3$,
\newauthor
  C. Adami$^{9}$, D. Alloin$^{1}$, A. Alshino$^{6}$, S. Andreon$^{10}$,  M. Birkinshaw$^{11}$, M. Bremer$^{11}$,
   A. Detal$^{5}$,
\newauthor
 P.-A. Duc$^{1}$, G. Galaz$^{12}$,  L. Jones$^6$,  J.-P. Le F\`evre$^{8}$, O. Le F\`evre$^{9}$, D. Maccagni$^{2}$,
 A.  Mazure$^{9}$,
\newauthor
H. Quintana$^{12}$, H.~J.~A.~R\"ottgering$^{13}$, P.-G.
Sprimont$^{5}$, C. Tasse$^{13}$,
\newauthor
G.~Trinchieri$^2$,  J.P. Willis$^{14}$\\
$^1$ Laboratoire AIM, CEA/DSM - CNRS - Université Paris Diderot,
DAPNIA/Service d'Astrophysique, Bât. 709, CEA-Saclay, F-91191
Gif-sur- Yvette Cédex, France\\
$^2$INAF, IASF Milano, via Bassini 15, I-20133 Milano, Italy\\
$^3$ESA, Villafranca del Castillo, Spain\\
$^4$Institue of Astronomy, Madingley Road, Cambridge CB3 0HA, UK\\
$^5$Institut d'Astrophysique et de G\'eophysique, Universit\'e de
Li\`ege, All\'ee du 6 Ao\^ut,
17, B5C, 4000 Sart Tilman, Belgium\\
$^6$School of Physics and Astronomy, University of Birmingham,
Edgbaston, Birmingham, B15 2TT, UK\\
$^7$Department of Physics and Astronomy, University of Leicester, Leicester LE1 7RH, UK\\
$^{8}$DAPNIA/SEDI CEA Saclay, 91191 Gif sur Yvette\\
$^{9}$ Laboratoire d'Astrophysique de Marseille, France\\
$^{10}$INAF, Osservatorio Astronomico di Brera, Milan, Italy.\\
$^{11}$Department of Physics,
University of Bristol, Tyndall Avenue, Bristol BS8 1TL, UK.\\
$^{12}$Departamento de Astronom{\'i}a y Astrof{\'i}sica,
Pontificia Universidad Cat{\'o}lica de Chile,
Santiago, Chile.\\
$^{13}$Leiden Observatory, P.O. Box 9513, 2300 RA Leiden, The
Netherlands.\\
$^{14}$Department of Physics and Astronomy, University of
Victoria, Elliot Building, 3800 Finnerty Road, Victoria, BC, V8P
1A1 Canada.\\}

\date{ }

\maketitle

\label{firstpage}

\begin{abstract}

Following the presentation of the XMM-LSS X-ray source detection
package by Pacaud et al., we provide the source lists for the
first 5.5 surveyed square degrees. The catalogues pertain to the
[0.5-2] and [2-10] keV bands and contain in total 3385 point-like
or extended sources above a detection likelihood of 15 in either
band. The agreement with deep logN-logS is excellent. The main
parameters considered are position, countrate, source extent with
associated likelihood values. A set of additional quantities such
as astrometric corrections and fluxes are further calculated while
errors on the position and countrate are deduced from simulations.
We describe  the construction of the band-merged catalogue
allowing rapid sub-sample selection and easy cross-correlation
with external multi-wavelength catalogues. A small optical CFHTLS
multi-band subset of   objects is associated wich each source
along with an X-ray/optical overlay. We make the full X-ray images
available in
 FITS  format. The data are available at CDS and, in a more extended
form, at the Milan XMM-LSS database.

\end{abstract}

\begin{keywords}
catalogues, surveys, X-rays: general
\end{keywords}

\section{Introduction}
\label{intro}

The XMM Large Scale Structure Survey (XMM-LSS)  has been designed
to provide a well defined statistical  sample of X-ray galaxy
clusters out to a redshift of unity, over a single large area,
suitable for cosmological studies \citep{pierre04}. This requires
the ability to detect and characterize faint extended sources, in
such a way as to control both the selection effects and the
contamination by spurious or misclassified pointlike sources. For
this purpose, we have developed a dedicated X-ray image processing
package, {\sc Xamin}, which is adapted to the complex
characteristics of the XMM  focal plane  \citep{pacaud06}. It is a
two-step procedure combining wavelet multi-resolution analysis and
maximum likelihood fits, both using Poisson statistics. The
package has been extensively tested and its parameters adjusted by
means of simulations, for the extended (clusters) and pointlike
(AGN) source populations; the latter representing some 95\% of the
X-ray sources at our sensitivity of $\sim 4~10^{-15}$ \flux\ in
the [0.5-2] keV band.

The first reports on cluster and AGN populations, based on the
{\sc Xamin} products and associated selection function, were
published by \citet{pierre06} and \citet{gandhi06} respectively.

The guaranteed time pointings (G fields in Table \ref{pointing})
were previously analysed with an independent, more traditional
pipeline analogous to the one used for the HELLAS2XMM survey
\citep{baldi02}, as described in \citet{chiappetti05}, providing
also the standard reference for the Milan  XMM-LSS database.

The XMM-LSS survey, located around $2^h30^m$~ $-5\deg$ is
associated in the optical with the Wide Synoptic component of the
Canada France Hawaii Telescope Legacy
Survey\footnote{http://cfht.hawaii.edu/Science/CFHLS/} (CFHTLS -
W1).

In this paper, we present the source lists obtained for the first
45 XMM-LSS pointings processed by {\sc Xamin} along with the
relevant information on the catalogues (Sec. \ref{slist}) as well
as the X-ray images (Sec. \ref{xray}). Sec. \ref{optdata}
describes the associated optical data that we  make publicly
available. The online facilities and plans for future releases are
presented in the two last sections.

\section{The X-ray source lists}
\label{slist}
\subsection{List of available pointings}
\label{pointings}

\begin{figure}
   \centering
  \includegraphics[width=8cm]{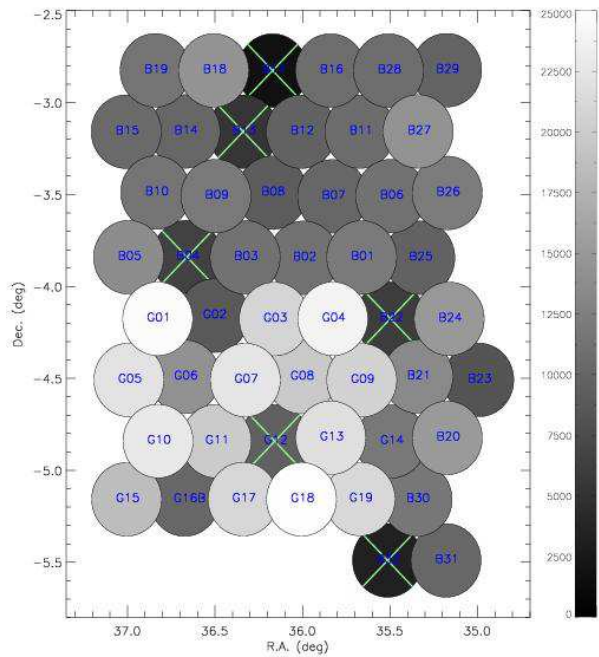}
\caption{Lay-out of the Guaranteed Time, AO-1 and AO-2 pointings.
The grey scale indicates the effective exposure times. Pointings
not included in the present release (see Table \ref{pointing}) are
flagged by a cross. The total resulting geometrical area, taking a
radius of $13'$ for the pointing useful radius (Sec. \ref{thresh})
amounts to 5.5 \dd.}
         \label{map}
\end{figure}

The XMM-LSS pointings pertaining to the Guaranteed Time,  AO-1 and
AO-2 periods are listed in Table \ref{pointing} and displayed in
Fig. \ref{map}. The raw X-ray observations (ODFs) were reduced
using the standard XMM Science Analysis System (XMMSAS; version
v6.1) tasks {\tt emchain} and {\tt epchain} for the MOS and pn
detectors respectively. High background periods, related to soft
protons, were excluded from the event lists following the
procedure outlined by \citet{Pratt02}.The resulting light
curves were visually inspected. After this operation, a number of
pointings appeared to be not usable for the purpose of the XMM-LSS
survey (too   low effective exposure times) and, consequently, are
not included in the present version of the catalogue. The fields
in question  were re-observed during AO-5 (July 2006 - January
2007) and   will be published in a subsequent data release.

\begin{table*}
\caption{The individual XMM-LSS pointings. Quoted exposures are
effective exposures computed after filtering high background
periods; pointings with too  low exposures (which have been
re-executed during the XMM AO-5) are indicated by a star and are
not included in the source list presented in this paper.
\newline Shifts are the astrometric corrections:
 $\Delta$RA/cos(Dec) = RA$_{XMM}$-RA$_{correct}$, $\Delta$Dec = Dec$_{XMM}$-Dec$_{correct}$ (see Sec.
\ref{position});   last column indicates whether the correction is
based on CFHTLS W1 data (1) , see Sec. \ref{assdata}, or   on the
USNO-A2 catalogue (2).
\newline
The letter G refers to the Li\`ege/Milan/Saclay Guaranteed Time
(nominal exposure time: 20 ks) and the letter B to the
observations performed during the Guest Observer phases (nominal
exposure time: 10 ks).
 }
 \label{pointing} \centering
\begin{tabular}{l l l l  l l l  l l l}
\hline \multicolumn{2}{c}{Field id} & RA (J2000) & Dec (J2000) &
\multicolumn{3}{c}{Exposure times (ks)} &
\multicolumn{3}{c}{Astrometric correction}\\
Internal & XMM ID & & & MOS1 & MOS2 & pn
& $\Delta$ RA ($''$) & $\Delta$ Dec ($''$) & Origin \\
\hline\hline
G01  & 112680101 & 02:27:25.4 & -04:11:06.4 &   24.6 & 25.3 & 21.4 & -1.60 & +0.53 & 1 \\
G02  & 112680201 & 02:25:54.2 & -04:09:05.6 &   10.1 &  9.7 &  6.7 & -1.07 & +0.00 & 1 \\
G03  & 112680301 & 02:24:45.6 & -04:11:00.8 &   21.8 & 21.7 & 17.3 & -1.07 & +1.07 & 1 \\
G04  & 109520101 & 02:23:25.3 & -04:11:07.6 &   25.5 & 25.8 & 19.5 & -1.60 & +0.53 & 1 \\
G05  & 112680401 & 02:28:05.1 & -04:31:08.1 &   23.5 & 23.9 & 12.5 & -1.07 & +1.07 & 1 \\
G06  & 112681301 & 02:26:34.4 & -04:29:00.8 &   16.4 & 16.6 & 10.5 & -0.53 & +0.53 & 1 \\
G07  & 112681001 & 02:25:25.3 & -04:31:07.1 &   22.5 & 25.1 & 18.6 & -0.53 & +0.00 & 1 \\
G08  & 112680501 & 02:23:54.6 & -04:29:00.1 &   21.2 & 21.3 & 15.9 & +0.00 & +1.07 & 1 \\
G09  & 109520601 & 02:22:45.2 & -04:31:11.1 &   22.5 & 22.7 & 16.4 & +0.00 & +0.53 & 1 \\
G10  & 109520201 & 02:27:25.4 & -04:51:04.4 &   24.7 & 24.6 & 18.5 & -1.07 & +1.07 & 1 \\
G11  & 109520301 & 02:26:05.1 & -04:51:06.1 &   21.7 & 21.8 & 16.1 & -1.06 & +0.53 & 1 \\
G12*  & 109520401 & 02:24:45.4 & -04:51:11.2 &   \multicolumn{3}{l}{not used} &-0.54 & +0.53 & 1 \\
G13  & 109520501 & 02:23:13.1 & -04:49:03.1 &   23.6 & 23.9 & 17.8 & -2.67 & -0.53 & 1 \\
G14  & 112680801 & 02:22:04.1 & -04:51:09.7 &   14.4 & 14.1 & ~8.3 & +1.06 & +0.53 & 1 \\
G15  & 111110101 & 02:27:54.1 & -05:09:02.3 &   20.8 & 21.8 & 14.0 & -2.67 & +0.00 & 1 \\
G16a* & 111110201 & 02:26:34.2 & -05:09:03.1 &   \multicolumn{3}{l}{not used} &   -1.60 & +0.53 & 1 \\
G16b & 111110701 & 02:26:35.2 & -05:08:46.6 &   11.9 & 11.9 & 11.5 & +0.00 & +0.53 & 1 \\
G17  & 111110301 & 02:25:14.3 & -05:09:08.4 &   22.4 & 22.2 & 17.5 & -2.67 & -0.53 & 1 \\
G18  & 111110401 & 02:23:54.1 & -05:09:09.7 &   27.7 & 28.0 & 19.2 & -2.67 & +0.00 & 1 \\
G19  & 111110501 & 02:22:34.0 & -05:09:02.1 &   23.2 & 23.8 & 16.8 & -2.67 & +0.00 & 1 \\
B01  & 037980101 & 02:22:45.5 & -03:50:58.8 &   14.1 & 14.2 & ~8.3 & -2.13 & +0.00 & 1 \\
B02  & 037980201 & 02:24:05.6 & -03:51:00.0 &   13.2 & 13.2 & ~7.8 & -2.13 & +0.00 & 1 \\
B03  & 037980301 & 02:25:25.7 & -03:50:59.2 &   13.3 & 13.0 & ~7.9 & -1.07 & +0.53 & 1 \\
B04*  & 037980401 & 02:26:45.4 & -03:51:00.1 &   \multicolumn{3}{l}{not used} &   -1.07 & +0.53 & 1 \\
B05  & 037980501 & 02:28:05.4 & -03:51:00.5 &   15.7 & 15.7 & 10.5 & -0.53 & +1.06 & 1 \\
B06  & 037980601 & 02:22:05.6 & -03:31:00.2 &   13.2 & 13.2 & ~7.7 & -1.60 & +0.53 & 2 \\
B07  & 037980701 & 02:23:25.7 & -03:30:56.7 &   12.3 & 12.3 & ~6.9 & -3.73 & +0.00 & 2 \\
B08  & 037980801 & 02:24:34.3 & -03:29:05.1 &   10.6 & 11.5 & ~6.3 & -0.53 & +1.07 & 2 \\
B09  & 037980901 & 02:26:05.4 & -03:31:01.1 &   13.9 & 13.8 & ~8.9 & -0.54 & +0.53 & 2 \\
B10  & 037981001 & 02:27:14.2 & -03:28:58.7 &   13.0 & 13.2 & ~8.6 & -1.07 & +2.13 & 2 \\
B11  & 037981101 & 02:22:34.2 & -03:09:02.5 &   12.4 & 12.3 & ~7.9 & -3.20 & +1.07 & 2 \\
B12  & 037981201 & 02:23:54.4 & -03:09:04.1 &   11.4 & 11.3 & ~7.0 & -3.20 & +0.00 & 2 \\
B13*  & 037981301 & 02:25:14.4 & -03:08:57.4 &   \multicolumn{3}{l}{not used} &   -0.53 & +1.60 & 2 \\
B14  & 037981401 & 02:26:34.4 & -03:08:57.6 &   12.6 & 12.5 & ~6.6 & -0.53 & +3.73 & 2 \\
B15  & 037981501 & 02:27:54.6 & -03:08:59.3 &   11.4 & 11.5 & ~8.4 & -2.13 & +1.07 & 2 \\
B16  & 037981601 & 02:23:14.4 & -02:48:56.3 &   12.3 & 12.7 & ~7.9 & -1.60 & +2.13 & 2 \\
B17*  & 037981701 & 02:24:34.8 & -02:48:50.0 &   \multicolumn{3}{l}{not used} &   -2.67 & +3.73 & 2 \\
B18  & 037981801 & 02:25:55.0 & -02:48:49.1 &   16.0 & 16.2 & 11.4 & -1.07 & +2.13 & 2 \\
B19  & 037981901 & 02:27:14.9 & -02:48:49.5 &   13.1 & 13.1 & ~8.7 & -1.07 & +2.13 & 2 \\
B20  & 037982001 & 02:20:34.8 & -04:48:46.3 &   16.1 & 16.8 & 11.7 & -1.60 & -0.53 & 1 \\
B21  & 037982101 & 02:21:14.9 & -04:28:45.9 &   15.6 & 15.6 & ~9.9 & -2.13 & +0.53 & 1 \\
B22*  & 037982201 & 02:22:05.5 & -04:11:03.9 &   \multicolumn{3}{l}{not used} &   +0.00 & +0.00 & 1 \\
B23  & 037982301 & 02:20:05.5 & -04:31:03.9 &   ~9.0 & ~9.6 & ~6.5 & -1.07 & +1.60 & 1 \\
B24  & 037982401 & 02:20:45.6 & -04:11:01.2 &   18.1 & 18.2 & 10.4 & -1.07 & +0.53 & 1 \\
B25  & 037982501 & 02:21:25.5 & -03:51:02.4 &   11.5 & 11.9 & ~6.8 & -1.07 & -0.53 & 1 \\
B26  & 037982601 & 02:20:34.8 & -03:28:50.4 &   14.7 & 14.6 & ~8.3 & -2.13 & +0.53 & 2 \\
B27  & 037982701 & 02:21:14.8 & -03:08:49.1 &   16.3 & 16.2 & 10.7 & -1.60 & +1.07 & 2 \\
B28  & 147110101 & 02:21:55.1 & -02:49:02.4 &   11.4 & 11.6 & ~9.1 & -1.60 & +0.00 & 2 \\
B29  & 147110201 & 02:20:35.1 & -02:49:00.1 &   10.6 & 10.6 & ~8.2 & -1.60 & +0.00 & 2 \\
B30  & 147111301 & 02:21:15.1 & -05:09:00.7 &   12.4 & 12.5 & ~9.9 & -1.07 & +0.53 & 1 \\
B31  & 147111401 & 02:20:36.3 & -05:28:59.5 &   10.8 & 11.0 & ~8.4 & -0.53 & +0.53 & 1 \\
B32*  & 147111501 & 02:21:56.0 & -05:28:56.4 &   \multicolumn{3}{l}{not used} &   -3.20 & +0.53 & 1 \\
\hline
\end{tabular}
\end{table*}

\subsection{Processing and {\sc Xamin} output source parameters}
\label{Xamin}

Photon images in different energy bands are then created with a
scale of 2\farcs5 pixel$^{-1}$.  Images from the three detectors
(pn, MOS1, M0S2), pertaining to the same band, are co-added. The
resulting image is in turn filtered in  wavelet
space\footnote{{\em \`a trou} algorithm, combined with a Poisson
noise model; see \citet{starck98}} to remove the noise at a given
significance level and subsequently scanned by a source detection
algorithm set to a low threshold to obtain a primary source list.
Detailed properties of each detected source are further assessed
from the individual photon images using {\sc Xamin}, a maximum
likelihood profile fitting procedure\footnote{the likelihood is
computed with respect to a flat image} designed for the XMM-LSS
survey. The specific goal of this second pass is to monitor in a
clean and systematic way the characterization of extended X-ray
sources and associated selection effects. The principle and
performances of {\sc Xamin}, for extended and point-like sources,
are described in detail by \cite{pacaud06} and we recall here the
main lines of the procedure. Basically, for each source, two
spatial emission models convolved by the XMM Point Spread Function
(PSF) are tested: ({\tt pnt}) a point source and ({\tt ext}) a
$\beta$-profile assuming a constant slope of 2/3. The main
parameters returned are the position and the fitted countrates
and, for model {\tt ext}, the best estimate of the core radius of
the $\beta$-model. Further, for each model, the
likelihood\footnote{The values actually provided are the natural
logarithm of the likelihoods but, to follow the widely spread
usage, we shall call them simply ``likelihood'' throughout the
paper. We refer to \citet[][Sec. 2.3.1]{pacaud06} for
clarification of this terminology issue. } of the source is
computed, as well as the likelihood of the extension for model
{\tt ext}. The fits are simultaneously performed on the pn and on
the MOS1+MOS2 sum images only requiring the source centre to be
the same in both images\footnote{In the case of model {\tt ext},
an additional fitting condition imposes that the core radii
inferred from the pn and MOS1+MOS2 images are the same}. The
procedure takes into account all main technical characteristics
such as: the blurring of the PSF and the vignetting as a function
of off-axis angle and photon energy, the gaps between the CCD of
the pn and MOS arrays, and the various background contributions.
Results of the {\tt ext} fit  allow us to define two classes of
extended objects in the {\sc Xamin} parameter space: the C1 and C2
classes (see Sec. \ref{class}). Table \ref{xaminpar} summarises
the output parameters of the pipeline. Table \ref{bkg}
provides the background level, averaged over all pointings
included in the present catalogue. The measurements were performed
on the X-ray images (obtained after removing bad time intervals)
by masking the detected source regions, for three ranges of
off-axis distance. We recall that the background consists of two
components: (1) the cosmic background which is affected by the
vignetting and (2) the particle background,  which is significant
at high energy and not affected by the vignetting. As a rule of
thumb, a point-like source detected with 15 net photons in any of
the 3 rings shows a signal to noise ratio of $\sim$  3.6 and 3.3
for the [0.5-2] and [2-10] keV bands respectively (when
considering a circular aperture having a radius of 10\arcsec ).

\begin{table}
\caption{{\sc Xamin} output parameters \citep{pacaud06}.
\newline See \citet{pacaud06} for the statistical definition of the
derived likelihood values
\newline Notes: $^a$  computed for both point-like and extended profile
fits, \newline $^b$ issued for each of the three EPIC detectors. }
\label{xaminpar}
\begin{tabular}{lll}
   \hline
   Parameter & Notes & Content\\
   \hline
   CUTRAD           &   & Size of the fitting box   \\
   EXP              & b & Mean exposure time in the box \\
   GAPFLAG          & b & Distance to nearest CCD gap   \\
   GAP\_NEIGHBOUR   &   & Distance to nearest detected neighbour \\ && in the fitting box \\
   EXT        &   & Best fit core radius      \\
   EXT\_LH          &   & Extension likelihood       \\
   DET\_LH          & a & Detection likelihood      \\
   X\_IMA,Y\_IMA     & a & Best fit position in pixel    \\
   RA,DEC           & a & Best fit sky coordinates  \\
   RATE\_MOS  & a & EPIC-MOS count rate       \\
   RATE\_PN   & a & EPIC-pn count rate        \\
   SCTS\_MOS        & a & Estimated source counts in MOS1+2 \\
   SCTS\_PN         & a & Estimated source counts in pn \\
   BG\_MAP\_MOS     & a & Background level in MOS1+2    \\
   BG\_MAP\_PN      & a & Background level in pn    \\
   PIX\_DEV         & a & Distance between input/output position\\
   N\_ITER          & a & Number of iterations in the fit   \\
   \hline
\end{tabular}
\end{table}

\begin{table}
\caption{Mean total background level (MOS1 + MS2 + pn) for the
pointings included in the present catalogue, as a function of
off-axis angle. Units are $10^{-6}$ counts/s/pixel for a pixel
size of 2.5\arcsec} \label{bkg}
\begin{tabular}{lll}
   \hline
   Off-axis & [0.5-2] keV &  [2-10] keV \\
   \hline
   0\arcmin - 5\arcmin          & 7.1  & 12.7   \\
    5\arcmin - 10\arcmin          & 6.3  & 12.4   \\
    10\arcmin - 13\arcmin          & 5.8  & 12.2   \\
   \hline
\end{tabular}
\end{table}

\subsection{Source list description}
\label{sourcedesc}

In this paper, we present the source lists for two bands, [0.5-2]
and [2-10] keV, named B and CD respectively. Tables \ref{catalog}
and \ref{catalogmerge} display the parameters we make available:
in addition to the {\sc Xamin} output, a number of parameters are
a posteriori calculated in order to facilitate the interpretation
of the data set. In its present state, {\sc Xamin} does not
perform error calculations. Mean statistical  errors were
estimated by means of extensive simulations. \citet{pacaud06}
presented a detailed account of uncertainty estimates for the
extended sources (count rate and core radius). In the present
paper, further error information is provided for the point source
population; we note that  only the first 2 digits are to be
considered significant for the count rate and for the core radius
as well as for the derived quantities.

\subsubsection{Thresholds}
\label{thresh}

\begin{figure}
 \includegraphics[width=8cm]{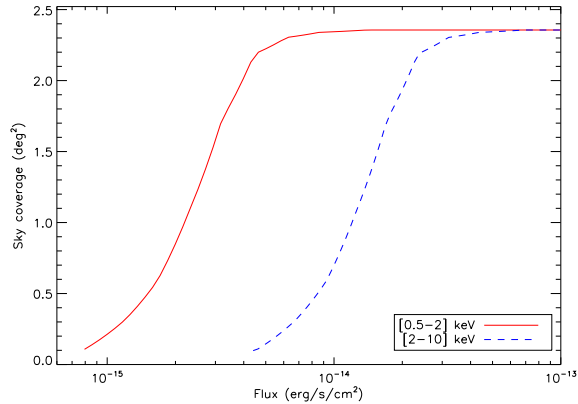}
\caption{  Sky coverage for the [0.5-2] and [2-10] keV bands
corresponding  to 27 pointings of 10 ks.  The curve in the soft
band is derived from extensive simulations of point sources and
includes only sources having a detection likelihood $>$ 15 within
off-axis distances less than 10$'$. The hard curve is scaled from
the soft one on the basis of equivalent signal to noise ratio,
using adequate background, vignetting and PSF characteristics.}
\label{skycover}
\end{figure}

{\sc Xamin} processes   sources only out to an off-axis distance
of 13$'$ by applying a detection mask centred on the mean optical
axis of the 3 telescopes, considering in this way only sources
that are visible by the 3 detectors. The total geometrical area of
the present catalogue corresponding to the validated pointings
indicated on Fig. \ref{map} amounts to 5.5 \dd. The present
version of the B and CD band catalogues contains the extended
sources identified as C1 and C2 (see Sec. \ref{class}) to which
are added sources having a point source detection likelihood
($LH$) greater than 15 (so-called {\it non-spurious}). As shown in
Fig. \ref{skycover}, this ensures a 90\% completeness limit of  4
$10^{-15}$ \flux\ in the B band within an off-axis distance of
10$'$ for the 10ks pointings (beyond 10$'$, the sensitivity falls
below 50\% of the value at the centre); in average, this
corresponds to 85 and 70 sources per 20 ks and 10 ks pointing
respectively, judged to be real within the inner 13$'$.

\subsubsection{Countrate and Flux}
\label{flux}

\begin{figure*}
\centerline{
 \includegraphics[width=6cm]{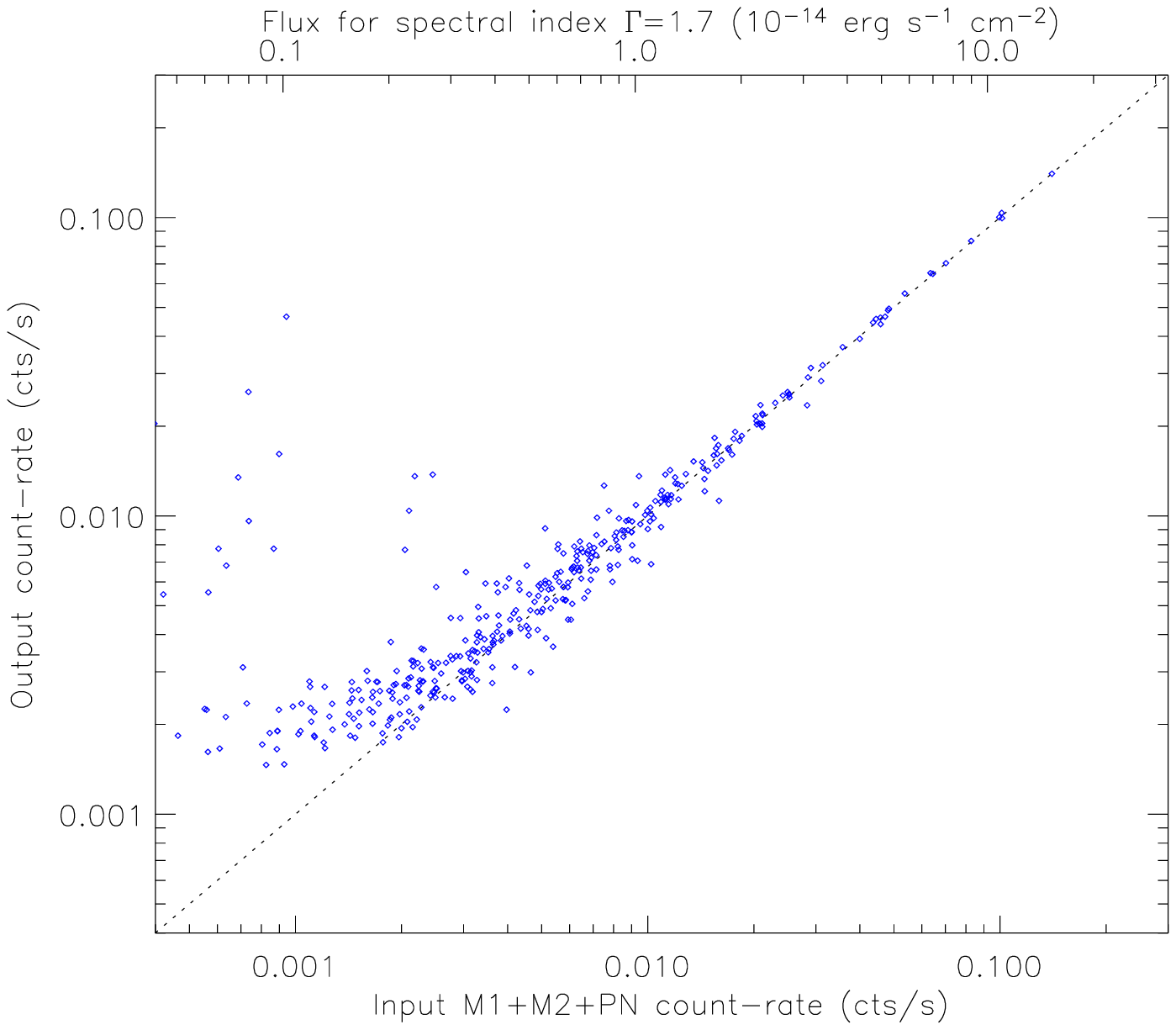}
 \includegraphics[width=6cm]{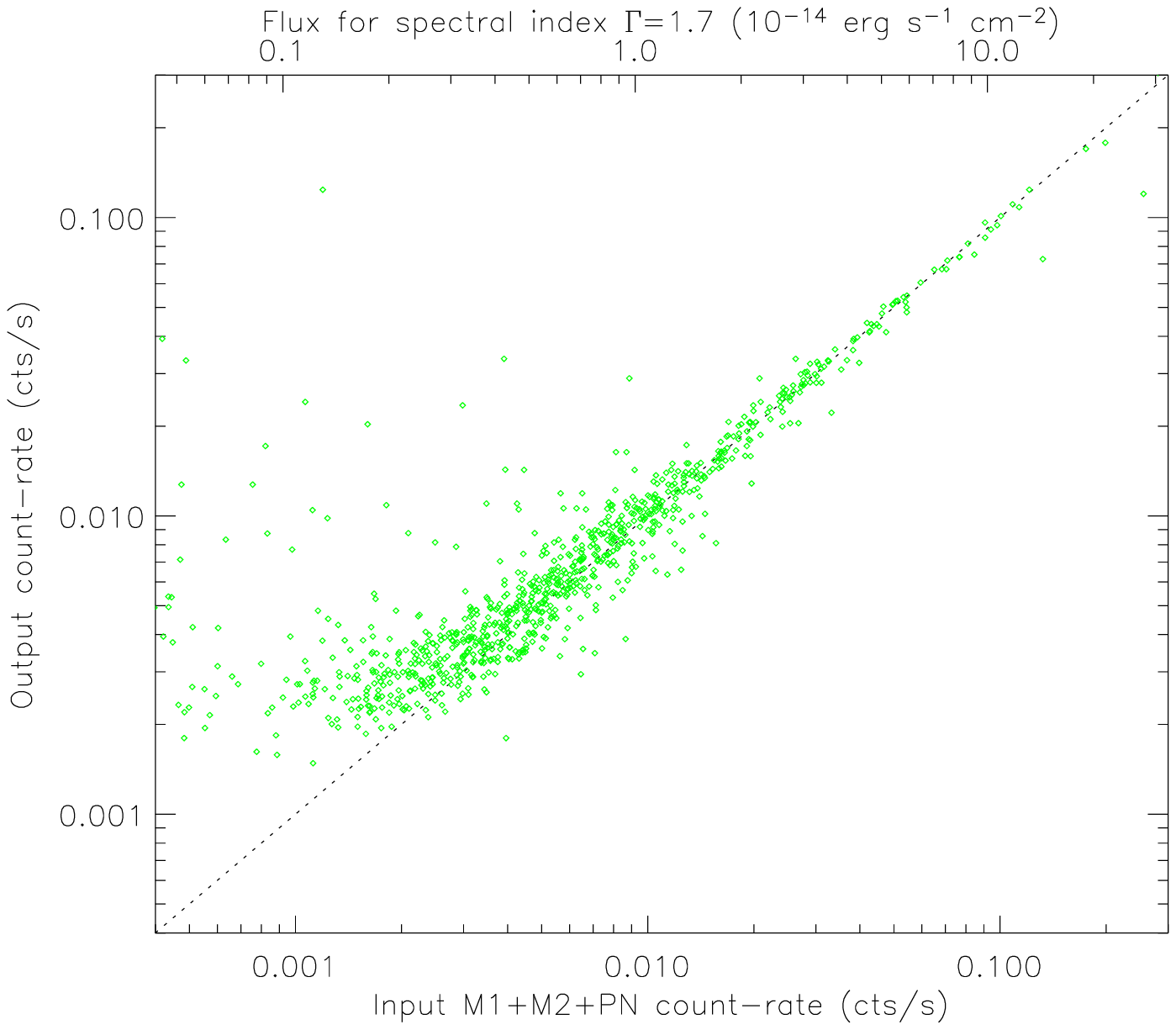}
  \includegraphics[width=6cm]{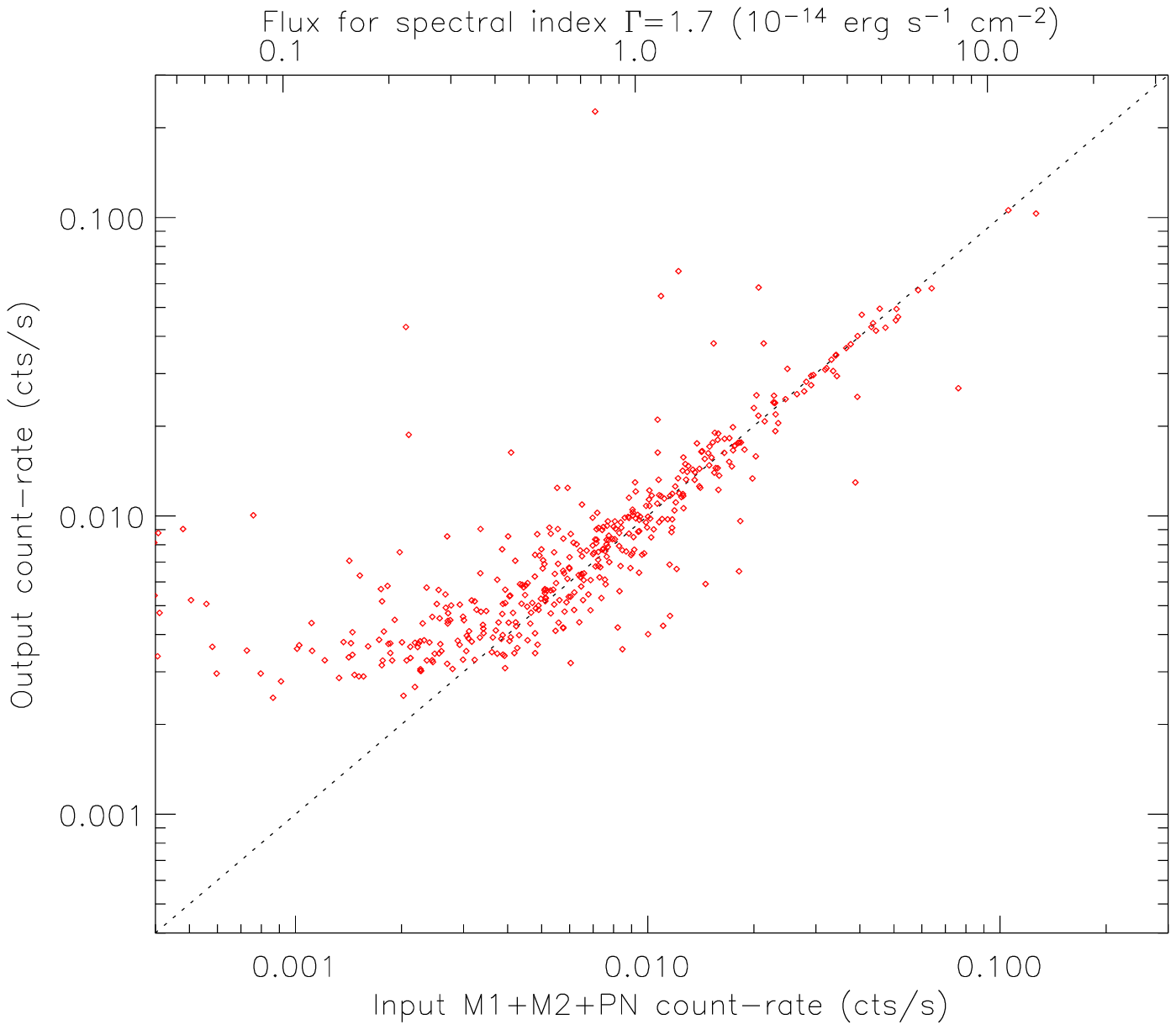}
 }
\caption{Photometric accuracy for three ranges of off-axis values
from 10 ks simulated images in the soft band: $0'-5'$, $5'-10'$,
$10'-13'$. ``Count-rate'' is the measured MOS1+MOS2+pn rate,
normalised to the on-axis value.}
         \label{photom}
\end{figure*}

{\sc Xamin} fits the MOS and pn count-rates
independently\footnote{while the source position and extent are
fixed to be the same for  the 3 detectors} (assuming that the two
MOS have the same response). We performed a number of point source
tests by simulating    a set of 30 images in which the logN-logS
distribution was injected. We display in Fig. \ref{photom} the
full range of simulations detailing the photometric accuracy out
to an off-axis distance of $13'$ \citep[][presented averaged
performances]{pacaud06}. The Eddington bias is obvious for faint
sources and increases with the off-axis distance; in particular
the photometry appears to be unreliable for a number of sources
detected with a total count-rate (pn+MOS1+MOS2) between 0.002 and
0.005 count/s  at off-axis angle $10'-13'$. These sources are
however detected with a likelihood greater than 15, indicating
that they are real. They are thus left in the public source
catalogue but their photometry should be handled with caution.
Mean values for the photometric bias and accuracy are summarized
in table \ref{tflux}.

\begin{table}
\caption{Mean photometric bias (b) and 1$\sigma$ error (e),
as a function of count-rate and off-axis distance for point
sources, for each of the B and CD bands. Values (in percentile)
are derived from 10 ks simulations considering sources having a
detection likelihood $> 15$ (cf Fig. \ref{photom}); below
count-rates of 0.003, the output locus is degenerate and the bias
is too large to estimate meaningful errors. In order to lower the
flux at which the bias arises, only sources having {\tt gap* $>12
''$}   (cf Table \ref{catalog}) are used for the error
calculation. The true count-rate is related to the observed
quantity by: $CR = CR_{obs}(100/ (b+100) \pm e/100$)}
 \label{tflux}
 \centering
\begin{tabular}{l l l}
 \hline \hline
 Band & B & CD \\
Countrate (count/s) &  b,~ e    &  b,~ e    \\
 \hline
 $0 <$off-axis$<5'$   & & \\
$0.003<CR<0.005$ & 9, 21& 9, 22\\
$0.005<CR<0.0075$ & 7, 15& 8, 16\\
 $0.007<CR<0.01$ & 6, 12 & 6, 12 \\
 $0.01<CR<0.02$ &4, 8 & 4, 8 \\
 $CR>0.02$ & 1, 4  &  1, 4\\  \hline
$5 <$off-axis$<10'$   & & \\
$0.003<CR<0.005$ & 10, 26& 11, 30\\
$0.005<CR<0.0075$ & 9, 19& 9, 21\\
 $0.007<CR<0.01$ & 7, 15 & 5, 11 \\
 $0.01<CR<0.02$ &5, 10 & 2, 5 \\
 $CR>0.02$ & 2, 5 & 1.3\\  \hline
 $10 <$off-axis$<13'$   & & \\
$0.003<CR<0.005$ & - & -\\
$0.005<CR<0.0075$ & 10, 24& 11, 29\\
 $0.007<CR<0.01$ & 9, 19 & 9, 21\\
 $0.01<CR<0.02$ &6, 13 & 7, 14 \\
 $CR>0.02$ & 3, 7 & 3, 7\\  \hline
 \hline
\end{tabular}
\end{table}

\begin{figure*}
\centerline{
\includegraphics[width=8cm]{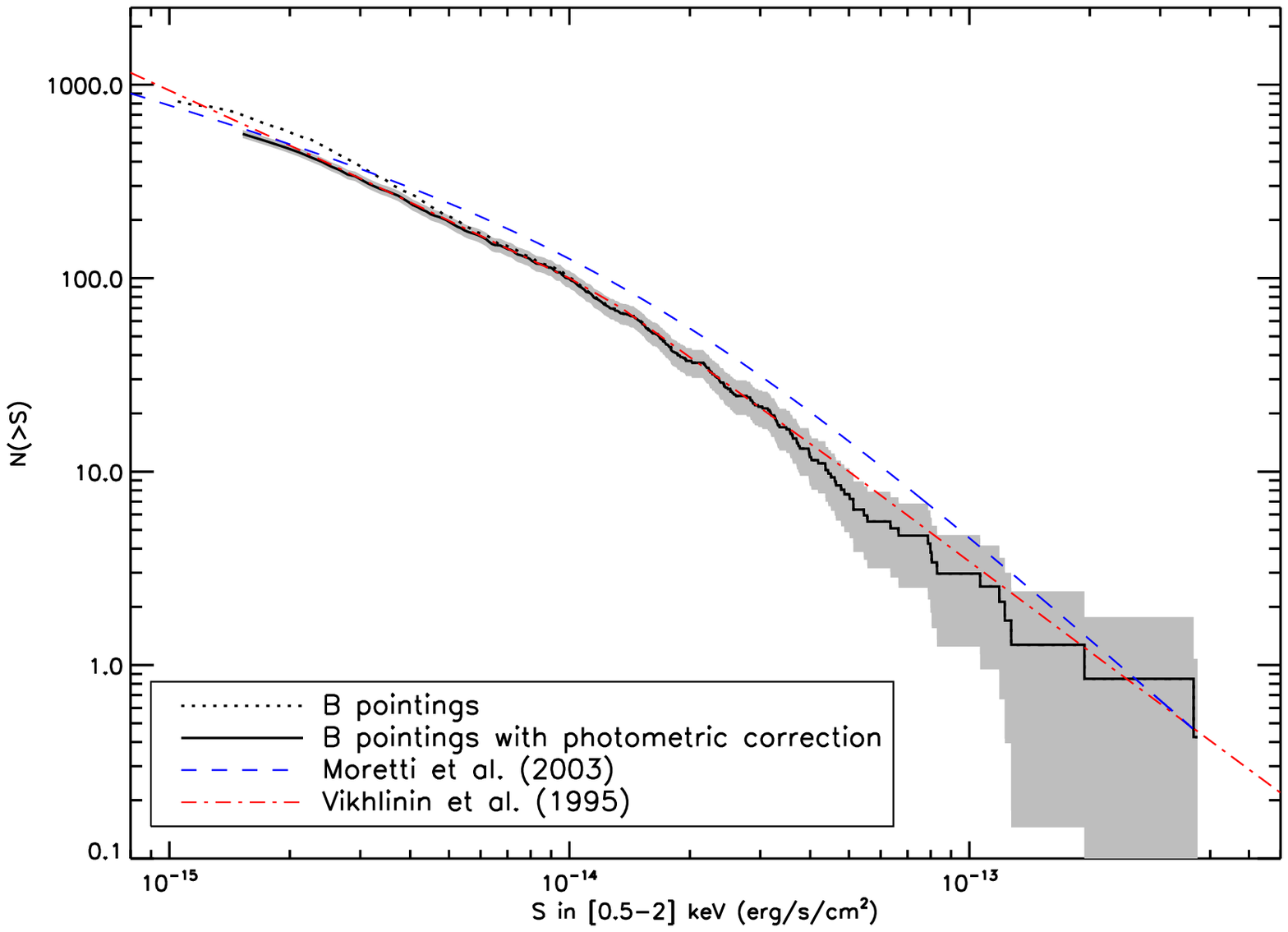}
\includegraphics[width=8cm]{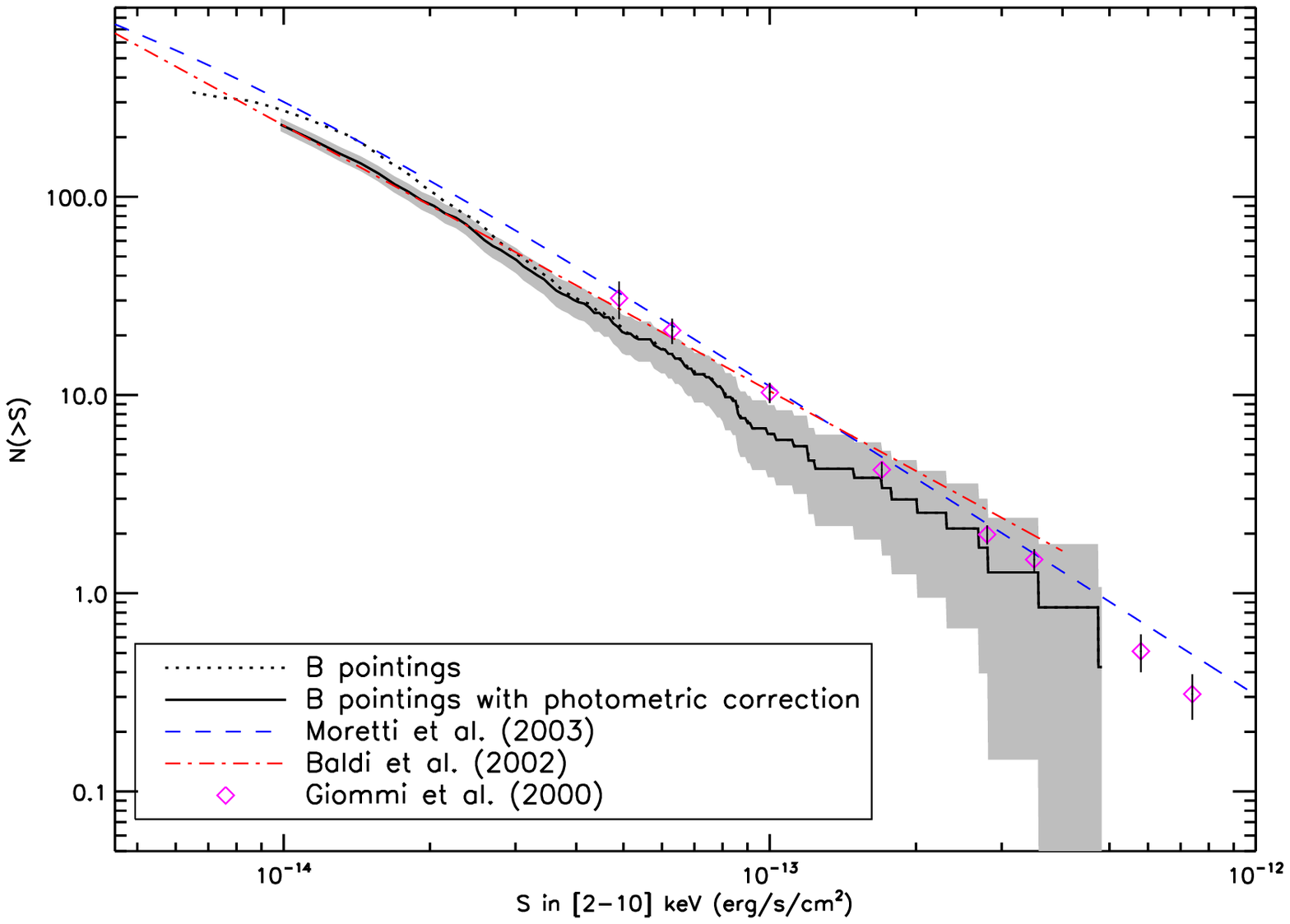}}
\caption{ The logN-logS distributions  for the soft (left) and
hard(right) bands, involving the 27 pointings with $\sim$ 10 ks.
Only sources having a detection likelihood $>$ 15 and an off-axis
distance less than 10$'$ are considered. The dotted line
corresponds to the raw counts and the solid line is corrected for
the Eddington bias \citep[as conspicuous in Fig. 7
of][]{pacaud06}. The dashed region indicates the 1$\sigma$
fluctuation level.}
\label{lognlogs}
\end{figure*}

Count-rate values are in turn converted into fluxes assuming a
standard power law spectrum (photon index of 1.7) and the mean $N_
{H}$ value of the region ($2.6 ~ 10^{20}$ cm$^{-2}$). The Energy
Conversion Factors are given in Table \ref{ecf}. { The observed
logN-logS distributions are presented in Fig. \ref{lognlogs}; they
are in good agreement with the \citet{vikhlinin95} (ROSAT),
\citet{moretti03} (ROSAT, Chandra and XMM), \citet{baldi02} (XMM)
and \citet{giommi00} (BeppoSAX) data points.

\begin{table}
\caption{The  Energy Conversion Factors for the individual EPIC
cameras and energy bands, stated in units of $10^{-12}$ \flux\ for
a rate of one count/s.  A photon-index power-law of 1.7 and a mean
$N_{H}$ value of $2.6~ 10^{20}$ cm$^{-2}$) are supposed. The two
MOS cameras are assumed to be identical. \label{ecf}}
\begin{tabular}{lll}
   \hline
   \hline
   Detector & B band & CD  band\\ \hline
   MOS & 5.0 & 23\\
   pn & 1.5 & 7.9 \\
   \hline \hline
\end{tabular}
\end{table}

Note that the resulting pn and MOS fluxes may be quite different
for some sources. In most of the cases this is due to the fact
that part of the source is occulted by a CCD gap in one of the
detectors to the point where the information is not recoverable.
Such cases can be identified from the gap-related columns (Tables
\ref{xaminpar} and \ref{catalog}) and  it is suggested to use the
fitted parameters obtained from the detectors on which the source
is not affected by a gap (if any). For the sake of simplicity, a
single mean flux value is provided
([FLUX(MOS)+FLUX(pn)]/2)\footnote{as the MOS countrate is about
1/3 of the pn countrate for cluster spectra, the averaged MOS flux
involves a number of photons comparable to that of the pn, thus
justifying the simple mean} along with column {\tt  fluxflag}
indicating the difference between the fluxes inferred from the
MOS1+MOS2 combination  and the pn (0: less than 20\%; 1:  between
20-50\%; 2: greater than 50\%).  For the two single-band
catalogues, the   sources are roughly equally distributed between
the three categories of flux quality. Fluxes assuming a thermal
spectrum as well as temperature and luminosity information for the
extended sources classified as C1 and C2, confirmed as clusters,
can be found in the XMM-LSS cluster
database\footnote{http://l3sdb.in2p3.fr:8080/l3sdb/}.

\subsubsection{Positional accuracy and  astrometric corrections}
\label{position}

The positional accuracy for the point source population was also
estimated from the simulations. Results are displayed in Fig.
\ref{fpos} and summarised in Table \ref{tpos}. We recall here that
for the point-like fit, as explained by \cite{pacaud06}, positions
are fixed to those of the first pass catalogue (Sextractor
detections on the wavelet filtered image).

\begin{figure*}

 \centerline{
 \includegraphics[width=7cm]{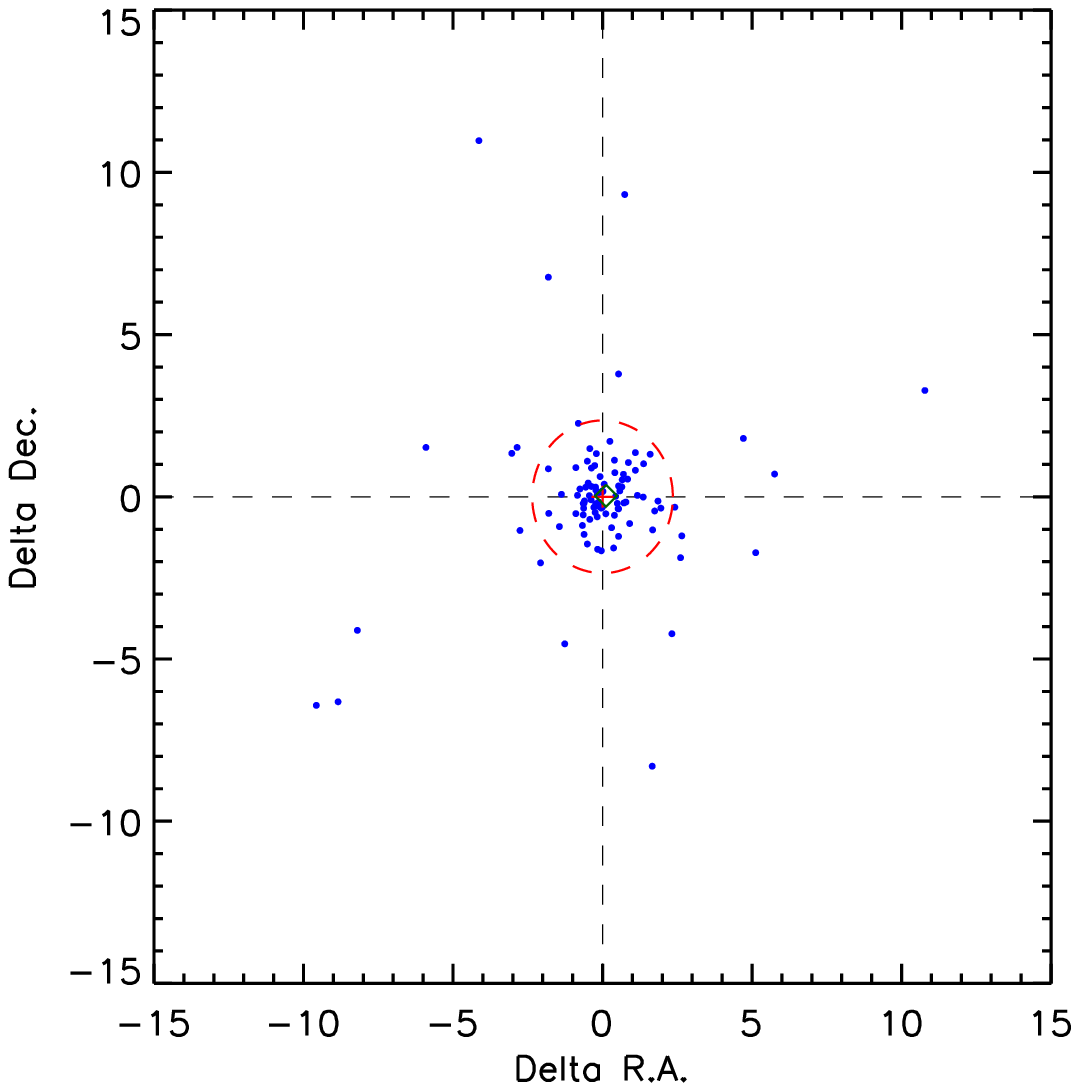}
 \includegraphics[width=7cm]{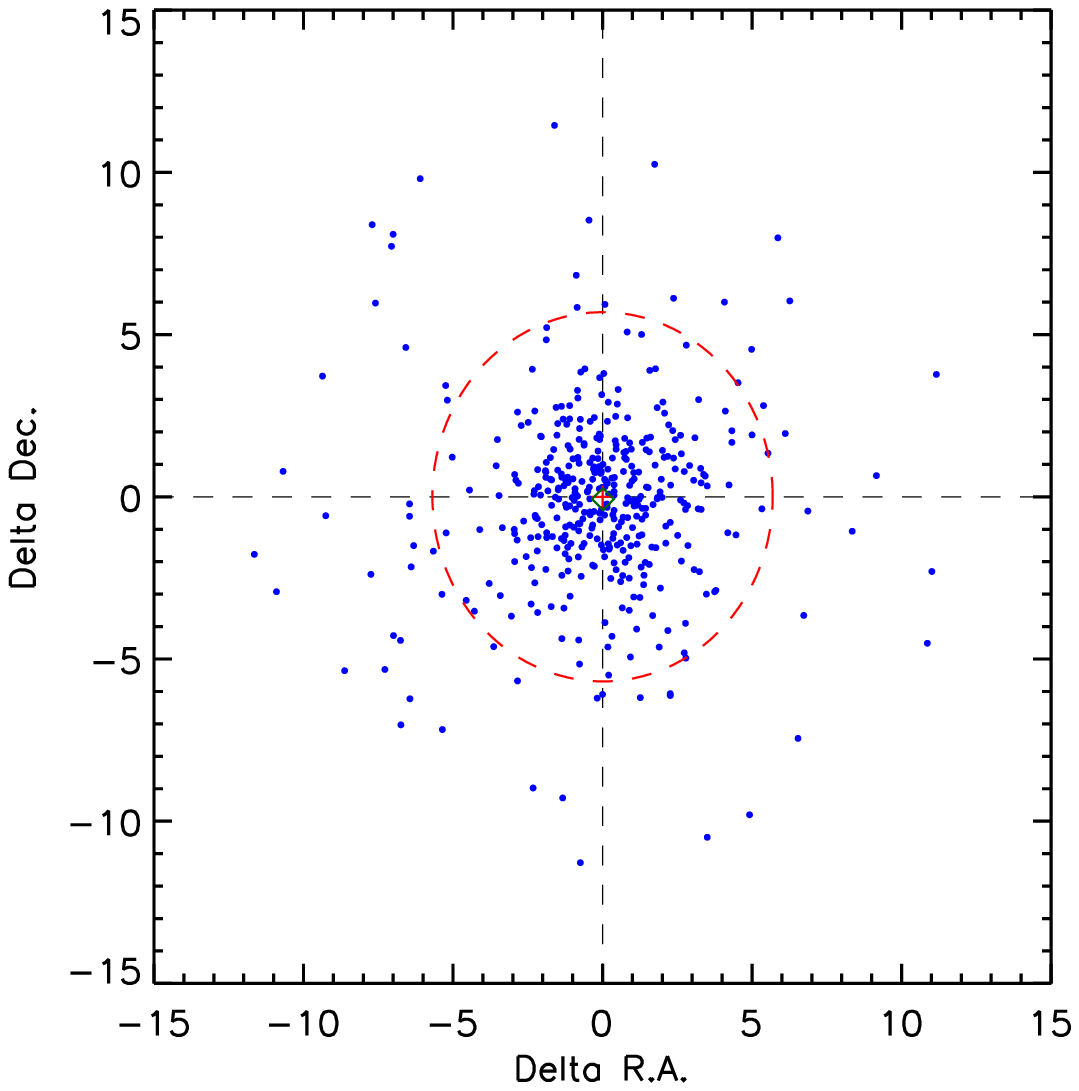}
 }
\caption{Example of positional accuracy diagrams  resulting from
the logN-logS point-source simulations of 10 ks exposures in the
soft band. Left panel: $input-output$ positions for the $0 <
$off-axis angle $  <5'$ and $CR > 0.01$ sources; Right panel:
$input-output$ positions for the $5' < $off-axis angle $ <10'$ and
$0.002<CR<0.005$ sources, where $CR$ is the measured MOS1+MOS2+pn
count-rate, normalised to the on-axis value. Axes are in unit of
arcsec. The circle indicates the 3$\sigma$ rejection radius used
to compute the mean positional error.}
         \label{fpos}
\end{figure*}

\begin{table}
\caption{ Positional accuracy (1$\sigma$ error on R.A. or Dec.)
for point sources derived from simulations of 10 ks pointings and
having a detection likelihood $> 15$ (Fig. \ref{fpos}). Values are
given for the B and CD bands, as a function of the  summed
measured count-rate: $CR$ = MOS1+MOS2+pn. No selection is applied
on the {\tt gap*} parameters (cf Table \ref{catalog})  but a
3$\sigma$ rejection is performed in the calculation of the errors.
Because of the strong Eddington bias for faint sources located
beyond $R>10'$ (Fig. \ref{photom}), no positional errors are
provided for output $CR$ below $0.002$ count/s.}
 \label{tpos}
 \centering
\begin{tabular}{l c c}
 \hline \hline
Band & B & {\bf CD} \\
  Countrate (count/s) &  Error   ($''$) & {\bf Error}  ($''$)\\
 \hline
 $0 <$off-axis$<5'$   & & \\
$0.001<CR<0.002$ & 2.0 & 2.0\\
 $0.002<CR<0.005$ & 1.7 & 1.7 \\
 $0.005<CR<0.01$ &1.3& 1.3 \\
 $CR>0.01$ & 0.8 & 0.8\\  \hline
 $5'<$off-axis$<10'$   & &\\
 $0.001<CR<0.002$ & 2.0 & 2.0\\
 $0.002<CR<0.005$ & 1.8 & 1.9\\
 $0.005<CR<0.01$ &1.5 & 1.5\\
 $CR>0.01$ & 1.0 & 1.0 \\ \hline
 $10'<$off-axis$<13'$   & \\
 $0.001<CR<0.002$  & - & -\\
 $0.002<CR<0.005$ & 1.9 & 2.0\\
 $0.005<CR<0.01$ & 1.6 & 1.7\\
 $CR>0.01 $& 1.2 &1.3 \\
 \hline
\end{tabular}
\end{table}

In parallel, in order to compensate for possible inaccuracies in
the XMM pointing positions, a global rigid astrometric correction
was estimated using the {\sc SAS} task {\sc eposcorr}. We
generated, for each XMM pointing, a reference X-ray list with all
``non spurious" point-like sources along with a reference optical
list containing all CFHTLS W1 objects within 6\arcsec~ of the
X-ray objects, being brighter than $i'=25$ and having a "good" or
"fair" chance probability $p<0.03$ as defined in Sec.
\ref{optstat}  (in the case where an X-ray source had more than
one optical counterpart candidates, we retained only the one with
the smallest chance probability). For the three topmost rows of
XMM pointings in Fig. \ref{map}, for which the CFHTLS data were not
yet processed, the optical list was made from the USNO-A2
catalogue. The reference files were fed into {\sc eposcorr} using
the parameter {\tt maxsig=2} to force removal of spurious matches.
The offsets computed by {\sc eposcorr} are plotted in Fig.
\ref{astro}. Offsets computed using USNO-A2 and CFHTLS objects are
usually compatible within errors (with the nominal {\sc eposcorr}
errors being larger in the USNO-A2 case). The offsets given in
Table \ref{pointing}   were applied to all coordinate sets for
each source in the database. The database column {\tt Xastrocorr}
indicates whether the origin of the astrometric correction is the
CFHTLS W1 or the USNO-A2 catalogue. Astrometrically corrected
positions are used  in the subsequent operations: removal of the
redundant sources, source naming and cross-identification with the
optical catalogue. Simulations of pairs of point sources separated
by 20\arcsec\ show that 67\% of the pairs are resolved for sources
having 30 photons each; all pairs being resolved when the sources
contain 500 photons.  As shown by \citet{pacaud06}, a
detection likelihood threshold of 15 eliminates almost all
spurious detections (the corresponding contamination level is
$\sim$ 1\%).

\subsubsection{Extended source classification}
\label{class}
 \citet{pacaud06} and \citet{pierre06} presented in
detail the criteria for defining galaxy cluster  candidates. The
selection is performed in the {\sc Xamin} output parameter space
obtained in the soft band. This band presents the highest signal
to noise ratio (S/N) at any redshift for typical cluster spectra
(as well as for galaxy thermal halos) thus ensuring the highest
completeness level for the extended source detection. The cluster
candidate sample consists of   two classes:

 - The C1 class is defined such that no point sources are
misclassified as extended and is described by {\tt extent}~$>
5\arcsec$, {\tt likelihood of extent}~$>33$ and {\tt likelihood of
detection\footnote{for the {\tt ext} fit}}~$>32$. Note that while
the C1 class is meant to be uncontaminated in terms of point-like
sources, it contains a few nearby galaxies whose X-ray emission is
unambiguously extended (Pacaud et  al.  submitted).

-  The C2 class is described by {\tt extent}~$>5\arcsec$ and
$15<$~{\tt likelihood of extent}~$<33$  (no condition on the
{\tt likelihood of detection}) and typically displays a
contamination rate of 50\%.

There are 73 C1 \& C2 objects  flagged in the catalogue, column
{\tt Bc1c2}. Note that, for the unique purpose of band merging
(see below), a similar classification has been applied in the hard
band. In this band, there are only 21 sources flagged as  C2 and 1
flagged as C1, which are not detected in the soft one. Subsequent
X-ray/optical inspection of these latter sources does not reveal,
as expected, any new cluster. They are mostly weak sources at
extreme off-axis angle, or unresolved doubles, or cluster X-ray
centroids displaced by more than 6\arcsec\ from the soft band
position.

\subsubsection{Band merging}
\label{merging}

Because X-ray colours serve as a useful piece of information for
numerous science studies, we provide, in addition to the
individual catalogues, the two-band merged catalogue. Band merging
is a delicate operation given the rather large XMM PSF (FWHM $\sim
6"$ at 1 keV on axis) and its variation with energy and off-axis
distance. It may also happen that a close pair is resolved in one
band only, because of too low a photon number in the other band.
In order to provide the community with an efficient and reliable
tool, the construction of the merged catalogue received special
attention and is described below.

As an X-ray source can be detected in one or two bands and, for
each band, is independently fitted by the extended and point
source models with the coordinates free, we adopt the following
merging procedure. For each band, a source is classified as
extended (E) if  it satisfies {\tt extent}~$>5\arcsec$ and
 {\tt likelihood of extent}~$>15$ (i.e. C1 or C2 class);
if not, it is classified as point-like (P). Then,  pointing per
pointing, we flag associations between the 2 bands within a search
radius of 6\arcsec . Note that we allow associations involving
spurious sources ($LH<15$) at most in one band. We kept the
information (rate, flux, etc.) about entries below this threshold
in the merged catalogue, since it could be more useful (e.g. for
upper limits) than no information at all, but we flag those cases
with {\tt Bspurious=1} or {\tt CDspurious=1}. Finally, for each
soft-hard couple in the merged catalogue, we define the {\tt best
band}, i.e. the band in which the detection likelihood of the
source is the highest and from which the coordinates are taken.
Details of the merging process are summarised in Table
\ref{tabmerging}.

Starting from 2980 non-spurious sources in the soft band  and 1255
non-spurious sources in the hard band, the resulting merged
catalogue contains 3385 sources: out of them   50.5\% are detected
as point-like in the soft band only,   36.3\% as point-like in
both bands, 10.2\% as point-like in the hard band only;  2.1\% are
candidate clusters of galaxies (soft band only). The remaining
0.8\% are mostly borderline cases (cf end of Sec. \ref{class}). We
note that in an extremely limited number of cases (4 couples of
entries) the merging process gives ambiguous results, i.e. a
detection in one band can be associated with two different
detections in the other band. The flagging and naming of such
cases is described in Sec. \ref{naming}.

\begin{table*}
\caption{Merging decisions and definition of the {\tt best band}.
\newline
A source is defined as extended (E) in a given band if it
satisfies {\tt extent}~$>5\arcsec$ and
 {\tt likelihood of extent}~$>15$ in this band. Otherwise it is defined as
point-like (P).
\newline For all sources, but the C1 clusters of galaxies, {\sl flux} is computed from the ECF factors given in Table
\ref{ecf} using the point source rates.
\newline For the C1 clusters, the fluxes are set to -1  as the reader is
addressed to the XMM-LSS cluster database (spatial and spectral
fitting providing accurate flux and luminosity measurements)
\newline
The numbers in [] indicate the number of spurious sources
encountered in the inter-band associations (the  counterpart in
and B or CD is flagged as spurious i.e. has $LH<15$)
 \label{tabmerging}}
\begin{tabular}{llllllll}
   \hline
     B band & CD  band&  {\tt best band} & Coordinates from & Flux in B & Flux in CD&  \# of sources\\
   \hline
    E & undetected&  B & B E\_fit &  -1 for C1, {\sl flux} for C2 &  -1 & 59   \\
    P & undetected& B & B P\_fit & {\sl flux} &  -1 & 1710    \\
    undetected& E & CD & CD E\_fit & -1 &  {\sl flux}  & 22  \\
    undetected& P & CD  & CD P\_ it & -1 &  {\sl flux} & 347    \\
    E & E & where {\tt detlik\_ext} is max & best band E\_fit & -1 for C1, {\sl flux} for C2 &  -1 for C1, {\sl flux} for C2 & 2\\
    E & P & where {\tt detlik\_ext} is max & best band E\_fit & -1 for C1, {\sl flux} for C2 &  -1 for C1, {\sl flux} for C2 & 12 [4 CD]\\
    P & E & where {\tt detlik\_pnt} is max & best band P\_fit & {\sl flux} & {\sl flux} & 5 \\
    P & P & where {\tt detlik\_pnt} is max & best band P\_fit & {\sl flux} & {\sl flux} & 1228 [36 B, 358 CD]\\
   \hline
\end{tabular}
\end{table*}

For {\it all} entries we provide in the database the distance
between the positions found by {\sc Xamin} in the two energy bands
(which in most cases could be used to solve the above
ambiguities). Such an inter-band distance, for all the cases
present in both bands, except for the few ambiguities, is within
2\arcsec~ in 35\% of the cases, within 4\arcsec~ in 79\% and above
5\arcsec~ only in 9\%. These percentages change to 39\%, 83\% and
7\% if we exclude detections flagged as spurious in one band. If
we compare the inter-band distance with the combination of the
position errors computed according to the prescription in Table
\ref{tpos}, we have that in 64\% of the cases the inter-band
distance is less than the $1\sigma$ error, in 1.4\% of the cases
is above $2\sigma$.

\subsubsection{Removal of redundant sources}
\label{redundant}

Finally, in the case of redundant objects detected in the regions
where the pointings overlap, we keep in the catalogues only the
detection pertaining to the pointing where the source is the
closest to the optical centre (cf columns {\tt Boffaxis,
CDoffaxis} in Table \ref{catalog}). Except for a few ambiguous
cases described in Section \ref{naming}, a value of 6\arcsec\ is
found to be the adequate search radius to identify redundant
detections. In this way, 280 doublets and a few triplets were
identified and reduced to a single source.

\subsubsection{Source naming}
\label{naming}

Objects in the merged catalogue are labelled following the
IAU-style convention, i.e. {\tt XLSS Jhhmmss.s-ddmmss}. The
coordinates used in assigning the name are the ones deduced after
the rigid astrometric correction, and chosen as official, i.e.
those for the {\sl best band}  (see Table \ref{tabmerging}).

In the individual band catalogues, the sources are assigned
supplementary names, following the same standard - {\tt XLSSx
JHHMMSS.S-DDMMSS} - where {\tt x} is {\tt B} or {\tt CD} and
stands for the B and CD bands respectively. In this case, the
coordinates correspond to the  extended (E) or point-like (P) fit
in the relevant band (Table \ref{tabmerging}). As a consequence of
the merging procedure (Sec.\ref{merging}), the {\tt XLSS}
coordinate designation will coincide with one of the {\tt XLSSB}
or {\tt XLSSCD} (i.e. for the {\tt best band}).

In a limited number of cases (8 entries), a  source in a band
happens to be associated  with two different objects in the other
band. These couples of catalogue entries are flagged by  a
non-zero value in column {\tt Xlink}, the registered value being a
pointer to the other ``ambiguous" entry in the couple. For those
cases, the ambiguity in the {\tt XLSS} name is resolved (when
necessary, i.e. in 6 out of 8 cases) by the addition of a suffix :
the two members of a couple will appear as {\tt XLSS
JHHMMSS.S-DDMMSSa} and {\tt XLSS JHHMMSS.S-DDMMSSb}.

\section{Associated data products}
\label{assdata} \subsection{X-ray images} \label{xray}

 For each pointing, we make available - via the XMM-LSS database in
 Milan-
the following images in FITS format. They are accessible as ``data
products" : for every source in the catalogue, a clickable link
points to the images of the relevant pointings.

- The B and CD  band  photon  images  for the 3 detectors (after
the event filtering).

- Exposure maps for the 3 detectors and the 2 bands. All images
have a pixel size of 2.5$''$. Note that the World Coordinate
System (WCS) of the X-ray images is the one generated by the {\sc
SAS}, therefore it does not take into account the astrometric
correction described in section \ref{position} and quoted in Table
\ref{pointing}. Consequently when overlaying X-ray source
positions exactly on  the X-ray images, one should use the
coordinates labelled as ``raw" in Table \ref{catalog}, although
this  does not make much difference for most of the sources, given
the pixel size.

\subsection{Optical data} \label{optdata}
\begin{figure*}
 \centerline{
 \includegraphics[width=9cm]{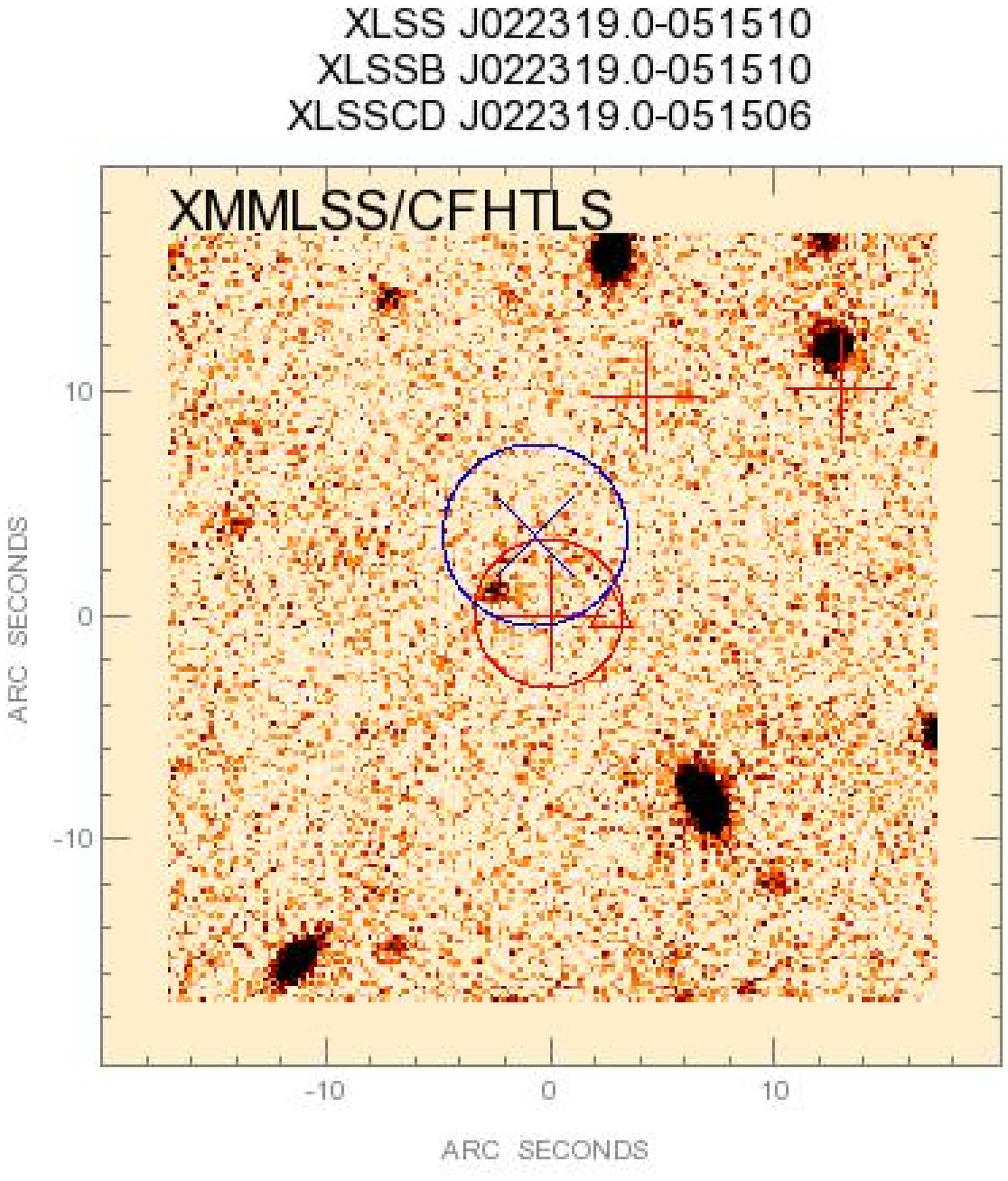}
 \includegraphics[width=9cm]{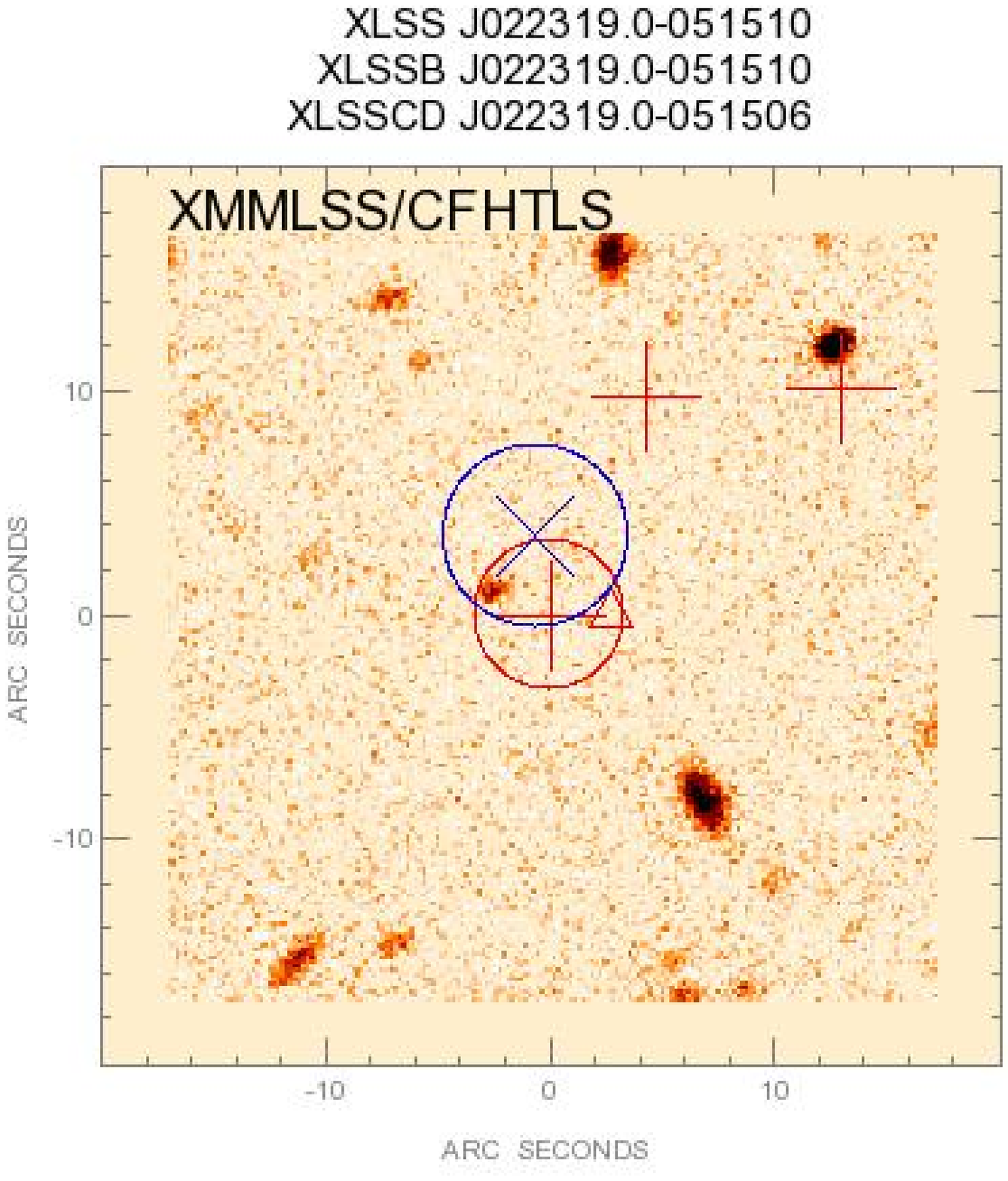}
 }
\caption {An example of postage stamp image centred on  source
XLSS J022319.0-051510. The  CFHT $i$ and $g$ $40'' \times 40''$
images are shown in left and right panels respectively. Soft
sources are indicated by a red +, and hard ones by a blue
$\times$; all with astrometrically corrected positions. The
original uncorrected position of the central source in the {\sl
best band} is indicated with a triangle of the corresponding
color. The 2 $\sigma$ radius mean error circles are indicated for
the central source according to the values of Table \ref{tpos}.
All sources within the $40'' \times 40''$ field having $LH>15$ are
plotted. North is up and East is left.} \label{stamp}
\end{figure*}

For each X-ray source of the  merged catalogue, we provide:

(i) the list of optical objects   within a radius of 6$''$ around
  each X-ray source, extracted from the CFHTLS catalogue\footnote{currently, from the
T0003 W1 field release by Terapix: http://terapix.iap.fr/}.  These
data are available in the XMM-LSS database in Milan only, through
the {\tt XLSSOPT} table. Queries on  X-ray lists are allowed and
return the $u^{*},g',r',i',z'$ magnitudes and further optical
information derived from the Terapix merged panchromatic
catalogues.

(ii) CFHTLS postage stamp images $40'' \times 40''$  in the $i$
and $g$ bands (PNG format); in the case where a CFHTLS image is
not yet processed, we provide a low exposure CFH12K  image when
available. An example is shown on Fig \ref{stamp}.

\subsection{A few statistics} \label{optstat}

 Out of our 3385 X-ray sources, 2208 have at least one optical
candidate closer than 6\arcsec~, 1071 fall in the area without
CFHTLS coverage\footnote{observed, but currently still under
processing} (corresponding to the topmost three rows of XMM
pointings in Fig. \ref{map}), while 106 have no optical
association. The total number of associated optical objects is
6275 (the radius used here is 6 arcsec irrespective of the fact
the object is extended or point-like). A total of 472 X-ray
sources have a single candidate within 6\arcsec~, 613 have two and
the rest has more. The number of optical counterparts brighter
than $i'=22.5$ is 1990 of which 55\%, 36\%, 14\%  are closer than
respectively 3\arcsec, 2\arcsec, 1\arcsec~ to an X-ray source
(detected in any band). If we consider only the soft (B) band
sources the numbers are 1826,
 56\%, 37\%, 14\% respectively, while for hard (CD) band sources
they are 1094,  59\%, 40\%, 17\%.

We further provide the distance $d$ between the X-ray and optical
positions as well as an estimate of the probability of chance
coincidence

\begin{displaymath}
  p = 1 - exp(-\pi~ n(<m)~  d^2 )
\end{displaymath}
where $n(<m)$ is the sky density of optical objects having an $i'$
magnitude brighter than the magnitude $m$ of the candidate
counterpart, computed from the full CFHTLS W1 catalogue. Defining
as ``good" or ``fair", respectively, the X-ray/optical matches
having $p<0.01$  and $0.01<p<0.03$ (as described by
\citet{chiappetti05}),   we find  945 good associations  and 637
fair ones. All the optical objects brighter than $i'=22.5$ and
within 1\arcsec~ and 87\% of those within 2\arcsec~ have a good
probability, while for a distance within 3\arcsec~ the percentage
lowers to 69\% for soft sources and 73\% for hard sources.

\section{Summary of online availability}
\label{online}

\begin{table*}
\caption{The XMM-LSS X-ray and optical data products. Individual
parameter availability is given in Tables \ref{catalog} and
\ref{catalogmerge}}
 \label{online1}
\begin{tabular}{llll}
   \hline
     \multicolumn{2}{l}{Data sets}&   Location &  Address\\
   \hline
    \multicolumn{2}{l}{Merged {\tt XLSS} catalogue, main parameters} &CDS& {\tt
    http://cdsweb.u-strasbg.fr/cgi-bin/qcat?J/MNRAS/vol/pag}\\\hline
 Single-band  catalogues: & {\tt XLSSB, XLSSCD } & Milan & {\tt http://cosmos.iasf-milano.inaf.it/$\sim$lssadmin/Website/LSS/Query} \\
 Merged  catalogue (all parameters): & {\tt XLSS} & &\\
 X-ray images & & & \\
 Optical catalogue : & {\tt XLSSOPT} & & \\
 Optical postage stamps & && \\

\hline
\end{tabular}
\end{table*}

Online data access is summarised in Table \ref{online1}. Namely,
we provide :
\begin{itemize}
\item  The raw {\sc Xamin} results in individual catalogues for
the B [0.5-2] keV and CD [2-10] keV bands. Only sources above a
detection likelihood of 15 are made available. Redundant sources
detected in overlapping regions of different pointings are
removed; data from the pointing where each object has the smallest
off-axis angle are retained in the catalogue. In addition, fluxes
assuming a power law spectrum are provided for each point-like
source.

\item  The  B-CD band merged catalogue assuming a correlation
radius of 6\arcsec . This required the definition of the {\sl best
band} from which a number of parameters, such as the position, are
taken and, hence, a selection of information from the single band
catalogues. Source counterparts in the other band are made
available even if they have a detection likelihood below 15 .

\item Optical:  panchromatic CFHTLS catalogue within 6\arcsec of
each source of the merged catalogue as well as i' and g' band
$40\arcsec \times 40\arcsec$ png images.

\end{itemize}

 The {\it main}
parameters (listed in Tables \ref{catalog} and \ref{catalogmerge})
of the merged X-ray    catalogue are available in electronic form
at the Centre de Donn\'ees de Strasbourg (CDS). The single-band
and band-merged catalogues with {\it all} columns, as well as the
associated data products (X-ray images and optical information)
are accessible, with fully interactive selection, through the
XMM-LSS database located in Milan and described  by
\citet{chiappetti05}. User login details can be found in the entry
web page.
 
\section{Future}
\label{future}

From now on, the XMM-LSS catalogue and associated data sets will
be regularly incremented following the receipt of new XMM and
CFHTLS pointings. In parallel, {\sc Xamin} is being upgraded with
the inclusion of information on the photon energy for the source
characterization, in such a way as to better discriminate between
AGN and cluster sources. When the new version is validated, we
foresee reprocessing the entire  data set and  making it publicly
available with the corresponding documentation.

\section*{Acknowledgments}

The results presented here are based on observations obtained with
XMM-Newton, an ESA science mission with instruments and
contributions directly funded by ESA Member States and NASA. The
cluster optical images were obtained with MegaPrime/MegaCam, a
joint project of CFHT and CEA/DAPNIA, at the Canada-France-Hawaii
Telescope (CFHT) which is operated by the National Research
Council (NRC) of Canada, the Institut National des Sciences de
l'Univers of the Centre National de la Recherche Scientifique
(CNRS) of France, and the University of Hawaii.  This work is
based in part on data products produced at TERAPIX and the
Canadian Astronomy Data Centre as part of the Canada-France-Hawaii
Telescope Legacy Survey, a collaborative project of NRC and CNRS.
AG acknowledges support from Centre National d'Etudes Spatiales.
The Italian members of the team acknowledge financial contribution
from contract ASI-INAF I/023/05/0. AD, OG, EG, PGS and JS
acknowledge support from the ESA PRODEX Programme "XMM-LSS", and
from the Belgian Federal Science Policy Office for their support.
HQ acknowledges partial support from the FONDAP Centro de
Astrofisica. PG is a Fellow of the Japan Society for the Promotion
of Science.


\begin{figure*}
   \centering
   \includegraphics[width=18cm,height=9cm]{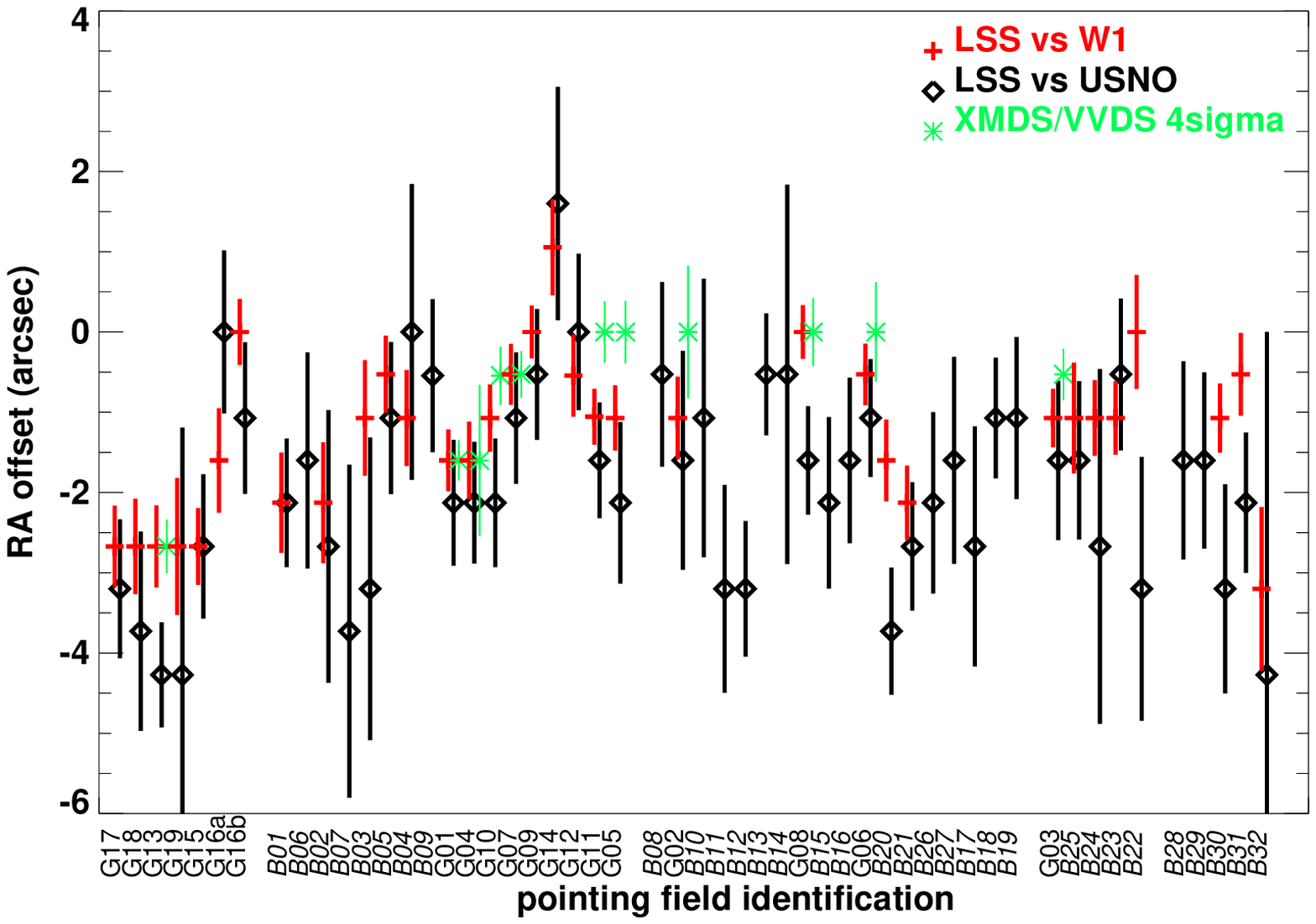}
   \includegraphics[width=18cm,height=9cm]{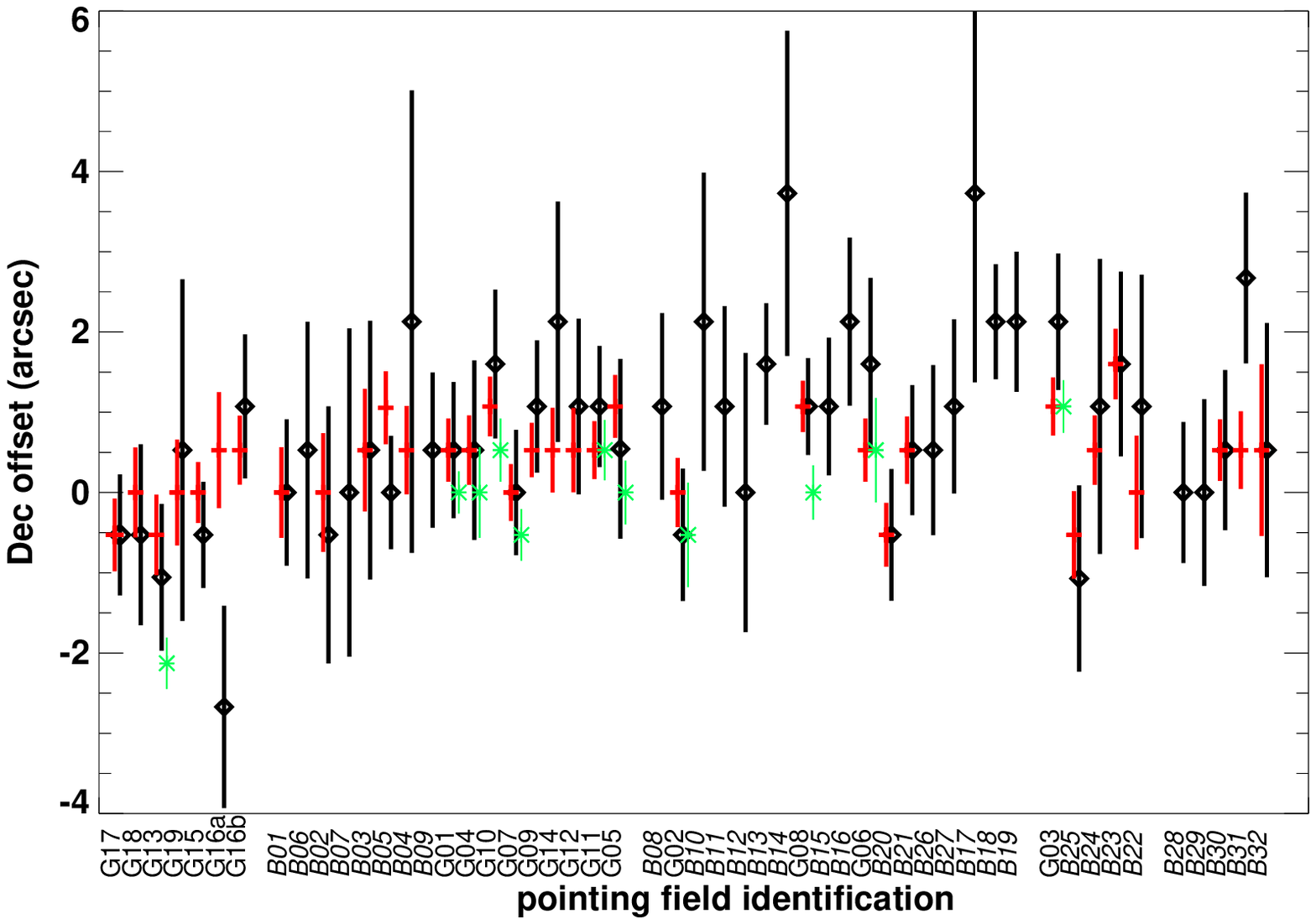}
  \caption{Astrometric correction offsets in RA (top panel) and Dec
(bottom panel), with nominal error bars computed by {\sc
eposcorr}. Diamonds   indicate the offsets computed using the
USNO-A2 catalogue, crosses those using CFHTLS W1 data, and
asterisks the offsets used in \citet{chiappetti05}, when
applicable. Different offsets for the same field are displaced
horizontally for clarity. The x-axis gives the XMM pointings in
chronological order of execution, with different observation
"seasons" separated by a blank space.}
         \label{astro}
\end{figure*}

\begin{table*}
\caption{List of parameters provided in the public XMM-LSS
catalogue. All are available at the XMM-LSS Milan database in
separate tables {\tt XLSSB} for the soft band and {\tt XLSSCD} for
the hard band (the column name has an appropriate prefix : when
there are two column names given, one with the prefix B and one
with the prefix CD, only the one applicable to the given band
appears in the relevant table but both may show up in the
band-merged table; column names without prefix are relevant to the
individual band only). The last four  columns indicate
respectively: (X)  whether a parameter is natively computed by
{\sc Xamin}; (m) whether a parameter is available also in the
band-merged table; (o) whether a parameter is present in the {\tt
XLSSOPT} table together with those described in Table
\ref{optical}; and (C) whether a parameter is present in the
catalogue stored at CDS. \label{catalog}}
\begin{tabular}{lllllll}
   \hline
   Column name & units & meaning and usage & X & m & o & C \\
   \hline
{\tt Bseq     or CDseq    } & --    & internal sequence number (unique)                     &   & X & X & X \\
{\tt Bcatname or CDcatname} & --    & IAU catalogue name {\tt XLSS{\it x} Jhhmmss.s-ddmmss}, {\tt{\it x}=B or CD} &   & X & X & X \\
{\tt Xseq}                & --    & pointer to merged entry see Table \ref{catalogmerge}  &   & X & X & X \\
{\tt Xcatname}            & --    & pointer to merged entry see Table \ref{catalogmerge}  &   & X & X & X \\
{\tt Xfield}              & --    & XMM pointing 1-32 for B01-B32 1001-1019 for G01-G19     &   & X &   &   \\
{\tt expm1}               & s     & MOS1 camera exposure in the band                      & X &   &   &   \\
{\tt expm2}               & s     & MOS2 camera exposure in the band                      & X &   &   &   \\
{\tt exppn}               & s     & pn   camera exposure in the band                      & X &   &   &   \\
{\tt gapm1}               & \arcsec & MOS1 distance to nearest gap                        & X &   &   &   \\
{\tt gapm2}               & \arcsec & MOS2 distance to nearest gap                        & X &   &   &   \\
{\tt gappn}               & \arcsec & pn   distance to nearest gap                        & X &   &   &   \\
{\tt Bnearest or CDnearest} & \arcsec & distance to nearest detected neighbour              & X & X &   &   \\
{\tt Bc1c2}               & 0|1|2   & 1 for class C1, 2 for C2, 0 for undefined                 &   & X & X & X \\
{\tt CDc1c2}              & 0|1|2   & 1 for class C1, 2 for C2, 0 for undefined                 &   &   &   &   \\
{\tt Bcorerad or CDcorerad} & \arcsec & core radius EXT (for extended sources)              & X & X &   & X \\
{\tt Bextlike or CDextlike} & --      & extension likelihood EXT\_LH                        & X & X &   & X \\
{\tt Bdetlik\_pnt or CDdetlik\_pnt} & --      & detection likelihood DET\_LH for pointlike fit      & X &   &   &   \\
{\tt Bdetlik\_ext or CDdetlik\_ext} & --      & detection likelihood DET\_LH for extended  fit      & X &   &   &   \\
{\tt Boffaxis or CDoffaxis} & \arcmin & off-axis angle                                      &   & X &   & X \\
{\tt Brawra\_pnt or CDrawra\_pnt}   & degrees & source RA  (not astrometrically corrected) for pointlike fit      & X &   &   &   \\
{\tt Brawdec\_pnt or CDrawdec\_pnt} & degrees & source Dec (not astrometrically corrected) for pointlike fit      & X &   &   &   \\
{\tt Brawra\_ext or CDrawra\_ext}   & degrees & source RA  (not astrometrically corrected) for extended  fit      & X &   &   &   \\
{\tt Brawdec\_ext or CDrawdec\_ext} & degrees & source Dec (not astrometrically corrected) for extended  fit      & X &   &   &   \\
{\tt Bra\_pnt or CDra\_pnt}         & degrees & source RA  (astrometrically corrected) for pointlike fit      & X &   &   &   \\
{\tt Bdec\_pnt or CDdec\_pnt}       & degrees & source Dec (astrometrically corrected) for pointlike fit      & X &   &   &   \\
{\tt Bra\_ext or CDra\_ext}         & degrees & source RA  (astrometrically corrected) for extended  fit      & X &   &   &   \\
{\tt Bdec\_ext or CDdec\_ext}       & degrees & source Dec (astrometrically corrected) for extended  fit      & X &   &   &   \\
{\tt Bposerr or CDposerr} & \arcsec & error on coordinates according to Table \ref{tpos} &   & X &   & X \\
{\tt Bratemos\_pnt or CDratemos\_pnt} & cts/s  & MOS count rate for pointlike fit          & X &   &   &   \\
{\tt Bratepn\_pnt or CDratepn\_pnt}   & cts/s  & pn  count rate for pointlike fit          & X &   &   &   \\
{\tt Bratemos\_ext or CDratemos\_ext} & cts/s  & MOS count rate for extended fit           & X &   &   &   \\
{\tt Bratepn\_ext or CDratepn\_ext}   & cts/s  & pn  count rate for extended fit           & X &   &   &   \\
{\tt countmos\_pnt  } & cts  & MOS number of counts for pointlike fit          & X &   &   &   \\
{\tt countpn\_pnt   } & cts  & pn  number of counts for pointlike fit          & X &   &   &   \\
{\tt countmos\_ext  } & cts  & MOS number of counts for extended fit           & X &   &   &   \\
{\tt countpn\_ext   } & cts  & pn  number of counts for extended fit           & X &   &   &   \\
{\tt bkgmos\_pnt    } & cts/pixel & MOS local background for pointlike fit          & X &   &   &   \\
{\tt bkgpn\_pnt     } & cts/pixel & pn  local background for pointlike fit          & X &   &   &   \\
{\tt bkgmos\_ext    } & cts/pixel & MOS local background for extended fit           & X &   &   &   \\
{\tt bkgpn\_ext     } & cts/pixel & pn  local background for extended fit           & X &   &   &   \\
{\tt Bflux or CDflux}                & erg/cm$^{2}$/s & source flux (undefined i.e. -1 for extended)  &   & X &   & X \\
{\tt Bfluxflag or CDfluxflag}        & 0 to 2         & 0 if MOS-pn difference $<20\%$, 1 between 20\%-50\%, 2 above 50\% &   & X &   & X \\
   \hline
\end{tabular}
\end{table*}

\begin{table*}
\caption{List of database parameters,  as Table  \ref{catalog},
but for the additional columns present only in the merged
catalogue table {\tt XLSS}.  When there are two column names
given, one with the prefix B and one with the prefix CD, they
relate to the given band, and both show up in the band-merged
table. Column names with the prefix X are relevant to merged
properties. } \label{catalogmerge}
\begin{tabular}{lllllll}
   \hline
   Column name & units & meaning and usage &X & m & o & C \\
   \hline
{\tt Xseq}                  & --      & Internal sequence number (unique)                 &   & X & X & X \\
{\tt Xcatname}              & --      & IAU catalogue name {\tt XLSS Jhhmmss.s-ddmmss{\it c}}, see Sec. \ref{naming}    &   & X & X & X \\
{\tt Bspurious and CDspurious} & 0 or 1  & set to 1 when soft/hard component has DET\_LH$<15$ &   & X &   &   \\
{\tt Bdetlike and CDdetlike}   & --      & detection likelihood EXT\_LH (pnt or ext according to source class)           & X & X &   & X \\
{\tt Xra}                   & degrees & source RA  (astrometrically corrected) (pnt or ext according to source class and in best band) &   & X & X & X \\
{\tt Xdec}                  & degrees & source Dec (astrometrically corrected) (pnt or ext according to source class and in best band) &   & X & X & X \\
{\tt Bra and CDra}           & degrees & source RA  (astrometrically corrected) (pnt or ext according to source class) &   & X & X & X \\
{\tt Bdec and CDdec}         & degrees & source Dec (astrometrically corrected) (pnt or ext according to source class) &   & X & X & X \\
{\tt Xbestband}             & 2 or 3  & band with highest likelihood : 2 for B, 3 for CD  &   & X &   &   \\
{\tt Xastrocorr}            & 1 or 2  & astometric correction from CFHTLS (1) or USNO (2) &   & X &   &   \\
{\tt Xmaxdist}              & \arcsec & distance between B and CD positions               &   & X &   &   \\
{\tt Xlink}                 & --      & pointer to Xseq of secondary association, see Sec. \ref{naming}  &   & X &   &   \\
{\tt Bratemos and CDratemos}   & cts/s   & MOS count rate (pnt or ext according to source class)  & X & X &   & X \\
{\tt Bratepn and CDratepn}     & cts/s   & pn  count rate (pnt or ext according to source class)  & X & X &   & X \\
   \hline
\end{tabular}
\end{table*}

\begin{table*}
\caption{ List of additional optical information presented as
columns in the {\tt XLSSOPT} table. The latter table also includes
the X-ray columns marked as such in Tables \ref{catalog} and
\ref{catalogmerge}. Therefore columns with the X, B or CD prefixes
refer to X-ray parameters, those with the O prefix to optical
data, and those without prefix to combined properties. The
arrangement of CFHTLS W1 fields is given at {\tt
http://terapix.iap.fr/cplt/oldSite/Descart/cfhtls/cfhtlswidemosaictargetW1.html}
 }
\label{optical}
\begin{tabular}{llll}
   \hline
   Column name & units & meaning and usage & Terapix\\
   \hline
{\tt Oseq}     & --        &  internal sequence number (unique)                                  & n/a        \\
{\tt Oid}      & --        &  original Terapix id in field                                       & {\tt id}   \\
{\tt Ofield}   & --        &  CFHTS field identification in form $\pm x \pm y$                   & n/a see caption \\
{\tt Ora}      & degrees   &  RA of the optical candidate                                        & {\tt ra}   \\
{\tt Odec}     & degrees   &  Declination of the optical candidate                               & {\tt dec}  \\
{\tt Oflag}    & --        &  binary flag combining 0/1 galaxy/star, 0/4 normal/masked, 0/8 normal/saturated  & {\tt flag} \\
{\tt Ou}       & magnitude &  $u^*$ magnitude                                                     & {\tt u} \\
{\tt Og}       & magnitude &  $g'$ magnitude                                                     & {\tt g} \\
{\tt Or\_}       & magnitude &  $r'$ magnitude                                                     & {\tt r} \\
{\tt Oi}       & magnitude &  $i'$ magnitude                                                     & {\tt i} \\
{\tt Oz}       & magnitude &  $z'$ magnitude                                                     & {\tt z} \\
{\tt Ou\_e}    & magnitude &  error on $u^*$ magnitude                                            & {\tt uerr} \\
{\tt Og\_e}    & magnitude &  error on $g'$ magnitude                                            & {\tt gerr} \\
{\tt Or\_e}    & magnitude &  error on $r'$ magnitude                                            & {\tt rerr} \\
{\tt Oi\_e}    & magnitude &  error on $i'$ magnitude                                            & {\tt ierr} \\
{\tt Oz\_e}    & magnitude &  error on $z'$ magnitude                                            & {\tt zerr} \\
{\tt distance} & \arcsec   &  distance from the X-ray corrected position to the optical position & n/a \\
{\tt prob}     & --        &  chance probability of X-ray to optical association (see text)      & n/a \\
   \hline
\end{tabular}
\end{table*}

\end{document}